\documentclass[aps, prb, twocolumn, amsmath, amssymb, nofootinbib, superscriptaddress]{revtex4-2}
\usepackage{hyperref}
\hypersetup{
    colorlinks = true,
    linkcolor = [rgb]{0.70, 0.13, 0.13},
    citecolor = [rgb]{0.25, 0.41, 0.88},
    urlcolor = [rgb]{0.25, 0.41, 0.88}
}
\usepackage{amsmath}
\usepackage{amsfonts}
\usepackage{amssymb}
\usepackage{pifont}
\usepackage{mathtools}
\usepackage{bm}
\usepackage{cases}
\usepackage{braket}
\usepackage[version=4]{mhchem}
\usepackage{graphicx}
\usepackage{cleveref}
\allowdisplaybreaks%

\begin{document}

\title{Gauge flux generations of weakly magnetized Dirac spin liquid in a kagom\'e lattice}
 
\author{Si-Yu Pan}
\affiliation{International Center for Quantum Materials, School of Physics, Peking University, Beijing 100871, China}
\author{Jiahao Yang}
\affiliation{International Center for Quantum Materials, School of Physics, Peking University, Beijing 100871, China}
\author{Gang v.~Chen}
\email{chenxray@pku.edu.cn}
\affiliation{International Center for Quantum Materials, School of Physics, Peking University, Beijing 100871, China}
\affiliation{Tsung-Dao Lee Institute, Shanghai Jiao Tong University, Shanghai 201210, China}
\affiliation{Collaborative Innovation Center of Quantum Matter, 100871, Beijing, China}

\begin{abstract}
Inspired by the recent progress on the Dirac spin liquid and the kagom\'{e} lattice 
antiferromagnets, we revisit the U(1) Dirac spin liquid on the kagom\'{e} lattice
and consider the response of this quantum state to the weak magnetic field by examining 
the matter-gauge coupling. Even though the system is in the strong Mott insulating regime, 
the Zeeman coupling could induce the internal U(1) gauge flux with the assistance of 
the Dzyaloshinskii-Moriya interaction. In addition to the perturbatively-induced non-uniform flux 
from the microscopic interactions, the system spontaneously generates 
the uniform U(1) gauge flux in a non-perturbative fashion to create the 
spinon Landau levels and thus gains the kinetic energy for the spinon matters. 
Renormalized mean-field theory 
is employed to validate these two flux generation mechanisms. 
The resulting state is argued to be an ordered antiferromagnet
with the in-plane magnetic order, and the gapless Goldstone mode behaves
like the gapless gauge boson and the spinons appear at higher energies.  
The dynamic properties of this antiferromagnet, and 
the implication for other matter-gauge-coupled systems are discussed. 
\end{abstract}

\maketitle  

\section{Introduction}
\label{sec:1}
  
Since the discovery of fractional quantum Hall effects, characterizing and identifying 
the exotic quantum phases of matter have attracted tremendous attention 
in modern condensed matter physics. Beautiful experiments were constructed 
to probe the emergent and fractionalized excitations as well as to reveal their novel properties. 
The success of fractional quantum Hall effects \cite{Tsui1982two,stormer1999fractional}, however, did not immediately carry over 
to other exotic matter. The prevailing interest in quantum spin liquids has not yet 
led to the firm experimental establishment. 
Nevertheless, in the long view, on top of the extensive theoretical progress, 
the field has identified many interesting quantum materials with unexpected 
physical properties and novel interactions. The mutual feedback between 
the theory and the experiments continues to inspire and push the frontier of the field.

The ongoing interest in the spin-1/2 kagom\'{e} lattice Heisenberg antiferromagnet 
(KHAFM) is accompanied by controversy on the candidate spin liquid ground states. 
Both the gapless U(1) spin liquid~\cite{ran2007projectedwavefunction} 
and the gapped $\mathbb{Z}_2$ spin liquid~\cite{lu2011z2,yan2011spinliquid,hao2009fermionic}
were proposed numerically and experimentally. 
Given the early numerical support for the U(1) Dirac spin liquid (DSL)  
ground state~\cite{ran2007projectedwavefunction} for the spin-1/2 KHAFM 
and the theoretical reasoning for the stability of the U(1) DSL on the 
frustrated lattices such as the kagom\'{e} lattice, 
it is natural to address the experimental connection in the real materials  
and establish some useful predictions for further experimental examination~\cite{hermele2008properties}. 
One interesting direction is to explore the consequences on the emergent degrees of freedom, 
such as the spinon matter and the gauge field, via the perturbation on the physical spin variables. 
The consequences on the emergent matter depend strongly on the nature of the ground state, 
and thus provide a useful testing of the actual ground state and the related excitations.  
In this work, we attempt to perform this analysis for the kagom\'{e} lattice antiferromagnet 
in the U(1) DSL state, and explain the manipulation of the spinon matter and the gauge fields.

The candidate materials for the KHAFM are in the strong Mott insulating regime, 
where the relevant physical degree of freedom is the localized electron spin. 
Therefore, the simplest available external perturbation that one can apply  
and control in experiments is the Zeeman coupling via the magnetic field. 
In addition, there exist the intrinsic perturbations to the Heisenberg model 
from the materials themselves, and the most noticeable ones are the antisymmetric 
Dzyaloshinskii-Moriya (DM) interactions~\cite{dzyaloshinsky1958thermodynamic, moriya1960anisotropic,lee2013proposal}. 
For the Cu$^{2+}$ ions under the consideration in the kagom\'{e} lattice antiferromagnets, 
the (symmetric) pseudo-dipole interactions are expected to 
be much weaker compared to the DM interaction and will not 
be considered in this work. This is because the pseudo-dipole interactions 
require one additional order of perturbation calculation in terms of 
the spin-orbit coupling for the microscopic Hubbard model deep in the Mott regime. 
With these available perturbations and the physical constraints, 
we plan to address the consequences on the kagom\'{e} lattice U(1) DSL 
from the behaviors of the fermionic spinons and the U(1) gauge fields  
and then provide the feedback to the experiments.

Apart from the relevance to the existing experiments and materials, understanding the role of 
the matter-gauge coupling for the lattice gauge theory is an interesting subject 
on its own. It has been shown that, 
a large number of gapless fermionic matter fields could stabilize 
the U(1) lattice gauge theory in 2D by suppressing the space-time monopole events 
via the matter-gauge coupling. This is also the underlying reason
for the possible existence of the U(1) DSL. 
More recently, on top of the usual confinement-deconfinement transition for the gauge theory,
numerical study of the $\mathbb{Z}_2$ lattice gauge theory with 
the fermionic matter found the spontaneous $\pi$-flux generation 
that converts the fermions from the gapless Fermi surface to 
the Dirac fermions in order to gain the kinetic energy~\cite{Gazit_2017}. 
This spontaneous flux generation is referred as the non-perturbative effect 
of the gauge-matter coupling in this work, and the induced flux via the DM interaction
and the Zeeman coupling is referred as the perturbative effect.

\begin{figure}[t]
\includegraphics[width=0.5\linewidth]{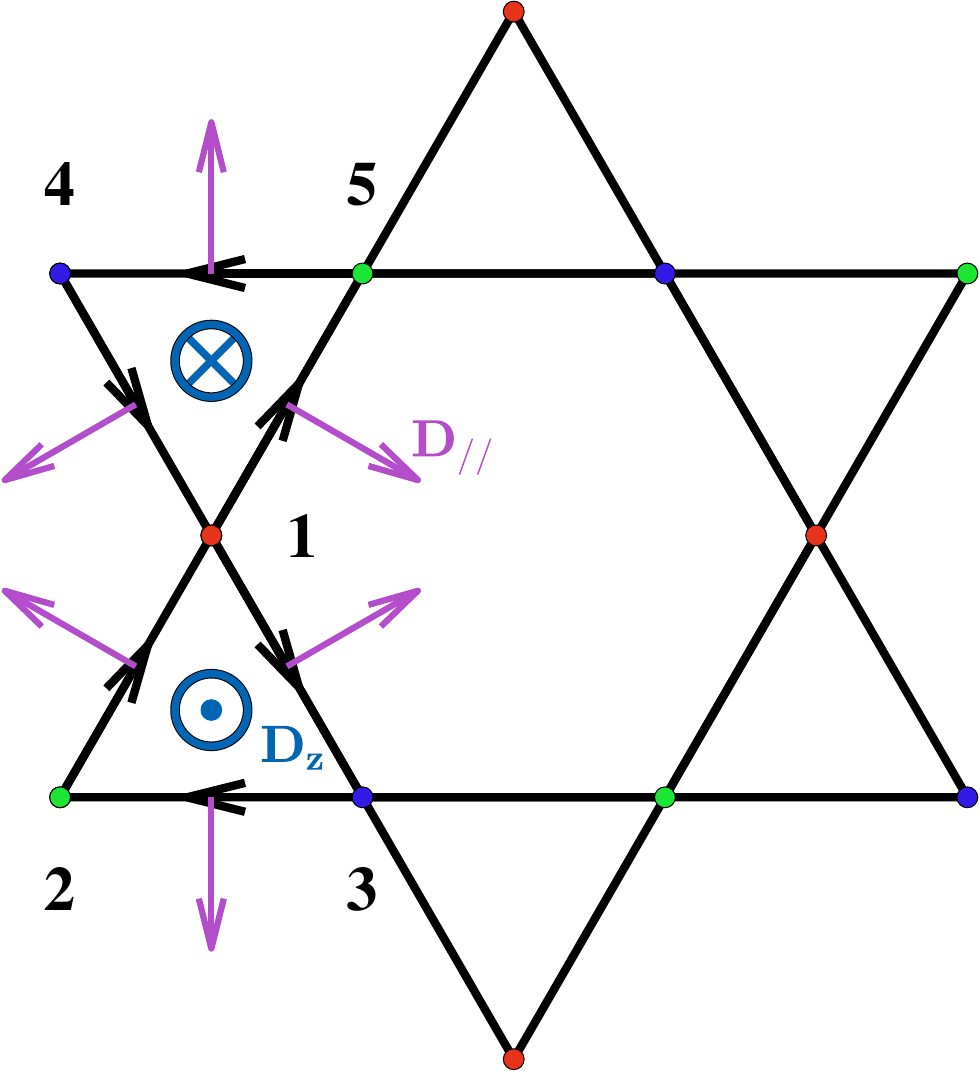}
\caption{Dzyaloshinskii-Moriya vectors on the kagom\'{e} lattice, 
with the in-plane ($D_\parallel$) and out-of-plane ($D_z$) components 
(the black arrow indicates the bond directionality (from site $j$ to $i$) 
that defines vector $\mathbf D_{ij}$.}
\label{fig:DM}
\end{figure}

To explain the key idea, we rely on the spin-1/2 spin model on the kagom\'{e} lattice, 
\begin{equation}
\label{eq:1}
    \begin{aligned}
    {\mathcal H} = \sum_{\braket{ij} } \left[ J {\mathbf S}_i \cdot {\mathbf S}_j  
                                + {\mathbf D}_{ij} \cdot ( {\mathbf S}_i \times {\mathbf S}_j ) \right] 
                                -  \sum_i {\mathbf B}\cdot {\mathbf S}_i ,
    \end{aligned}                             
\end{equation}
where the DM vector ${\bf D}_{ij} $ only has two components for each bond, 
i.e. the in-plane one ($D_{\parallel}$) and the out-of-plane one ($D_{z}$). 
In Fig.~\ref{fig:DM}, we depict the distribution of the DM vector on each bond 
of the kagom\'{e} lattice \cite{cepas2008quantum, elhajal2002symmetry, gao2020topological}.  
To be specific and simple, we focus on the out-of-plane one in this work. 
It was observed that, the U(1) gauge field fluctuation for the U(1) DSL ground state, 
that is related to the fluctuation of the scalar spin chirality 
$ ({\bf S}_i \times {\bf S}_j ) \cdot {\bf S}_k$~\cite{wen1989chiral},
is contained in the dynamic spin structure factor with the assistance of 
the DM interaction~\cite{lee2013proposal}. 
Based on this result, it was further observed that, 
the simple Zeeman coupling could induce an internal U(1) gauge flux 
distribution with the assistance of the DM interaction in a perturbative fashion~\cite{gao2020topological}. 
Since the spinon band structure and wavefunction are immediately modified with the modified background U(1) gauge field distribution, this observation leads to the interesting possibility 
for manipulating the properties of the spinons and the gauge fields 
for the U(1) DSL in the KHAFM and other similar contexts. 
Actually this question has not yet been studied for the U(1) DSL, 
and it is particularly useful to examine the U(1) DSL proposal 
for the kagom\'{e} spin liquid materials. 
On the other hand, the spontaneous flux generation introduced 
by the magnetization has been previously argued for the U(1) DSL on the kagom\'{e} lattice 
by Ref.~\onlinecite{ran2009spontaneous} in favor of the Landau level state (LL state), 
which is an imbalanced filling of the spinon Landau levels of the massless Dirac cones.
From the perspective of the spinons, the Zeeman coupling could create a chemical potential 
imbalance between different spin sectors, raising the energy of the U(1) DSL. 
This will make it a better choice to fill the magnetized Landau levels induced 
by the spontaneous uniform flux fluctuating from the U(1) DSL of KHAFM. 
Through the Zeeman coupling, this energetic advantage non-perturbatively 
stabilizes the system in the LL state.

This work combines the two flux generation mechanisms  
and addresses the consequences on the physical properties of the system.
We find that the Dirac cones in the U(1) DSL of KHAFM acquire mass 
through the combined effects of DM interaction and Zeeman coupling, 
rendering the Landau level state of the massive Dirac cones more stable 
with the finite out-of-plane magnetization. 
The chiralities of the massive Dirac cones near the Fermi surface, 
that are tuned by both DM interaction and magnetization, 
will determine the direction of the spontaneous uniform gauge flux. 
Furthermore, the magnitude of the out-of-plane magnetization 
directly controls the strength of this uniform flux. 
Using the duality argument, the massive Landau level state 
is an ordered antiferromagnet with the in-plane antiferromagnetic order~\cite{ran2009spontaneous,hermele2008properties}. 
The excitations of this state has three parts, gapless gauge photon at low energy,
regular spin-wave excitations, and the spinon continuum at higher energies, 
where the 
gauge photon is interpreted as the gapless Goldstone mode
of the spin waves~\cite{ran2009spontaneous}.

To deliver the results in a progressive manner, we first consider 
the perturbative flux generation from the DM interaction, 
and then incorporate the spontaneous flux generation mechanism 
based on the modified spinon bands of the perturbative flux generation. 
The remaining parts of the manuscript are organized as follows. 
In Sec.~\ref{sec:2}, we introduce the U(1) DSL for the kagom\'{e} lattice 
as the setting for our perturbative analysis. 
In Sec.~\ref{sec:3}, we explore the outcome of the perturbative gauge flux 
generation from the DM interaction on the spectroscopic properties 
of the spinon continuum and the wavefunction properties of the spinon bands. 
This includes the Berry curvature distribution and the spinon thermal Hall effects.
In Sec.~\ref{sec:4}, we study the massive Landau level state with spontaneous 
uniform flux based on the massive Dirac spin liquid of the kagom\'{e} lattice. 
We further compare the energy of the massive Landau level state 
and the massive DSL from the perspective of mean-field theory.
In Sec.~\ref{sec:5}, we apply the renormalized mean-field theory (RMFT) to 
validate the relationship between staggered flux, uniform flux, DM interaction strength, 
and magnetization within the mean-field framework.
The physical properties are then discussed with the results from the RMFT. 
Finally, in Sec.~\ref{sec:6}, we conclude with a discussion about 
the spinon Landau level state and the matter-gauge coupling in the lattice gauge theory.

\section{U(1) Dirac spin liquid as the parent state}
\label{sec:2}

To begin with, we sketch the early proposal by Y. Ran et al 
for the U(1) Dirac spin liquid ground state for the kagom\'{e} 
lattice Heisenberg model~\cite{ran2007projectedwavefunction}, 
which employs a fermionic parton construction to describe the Heisenberg part 
in Eq.~\eqref{eq:1}. 
In the fermionic parton language, the physical spin is expressed 
as $\mathbf S_i=\frac{1}{2}f^{\dagger}_{i\mu} {\boldsymbol \sigma}_{\mu\nu} f_{i\nu}$, 
where $f^{\dagger}_{i\mu}$ ($f^{}_{i\mu}$) is the spinon creation (annihilation) operator 
at the lattice site $i$ with the spin quantum number $\mu$, 
and ${\boldsymbol{\sigma}}$ is the Pauli matrix vector. 
This parton formulation enlarges the physical Hilbert space, and a local constraint,  
${\sum_{\sigma}f^\dagger_{i\sigma} f^{}_{i\sigma}=1}$ is imposed to 
return back to the physical Hilbert space. 
Using the spinon coherent state path integral and the standard Hubbard-Stratonovich transformation 
for the Heisenberg model, we obtain the quadratic action of the fermionic spinons with
\begin{equation}
    \begin{aligned}
    \mathcal S=& \int d\tau \Big[\sum_i f^{\dagger}_{i\sigma}\partial_\tau^{} f^{}_{i\sigma}
         +i\lambda^{}_i(f^\dagger_{i\sigma}f^{}_{i\sigma}-1) .\\
      +&\sum_{\langle ij \rangle }2J\left|\chi_{ij}\right|^2 
       +J(\chi_{ij}^{} f^\dagger_{j\sigma} f_{i\sigma}^{} +H.c.)       
      \Big],
    \end{aligned}
    \label{eq:2}
\end{equation}
where $\lambda_i$ is the Lagrangian multiplier to enforce the Hilbert space constraint, and 
$\chi_{ij}$ is the complex field for each bond introduced by the Hubbard-Stratonovich transformation.   
The mean-field states are specified by the $\{\chi_{ij} \}$ and $\{ \lambda_i \}$ configurations.  
A systematic classification of different mean-field spin liquid states can be performed based on 
the so-called projective symmetry group analysis. In Eq.~\eqref{eq:2}, the spinon pairing term is absent, 
and thus no $\mathbb{Z}_2$ spin liquid is considered in this choice.

\begin{figure}[t]
 \includegraphics[width=1\linewidth,trim=90 20 40 0, clip]{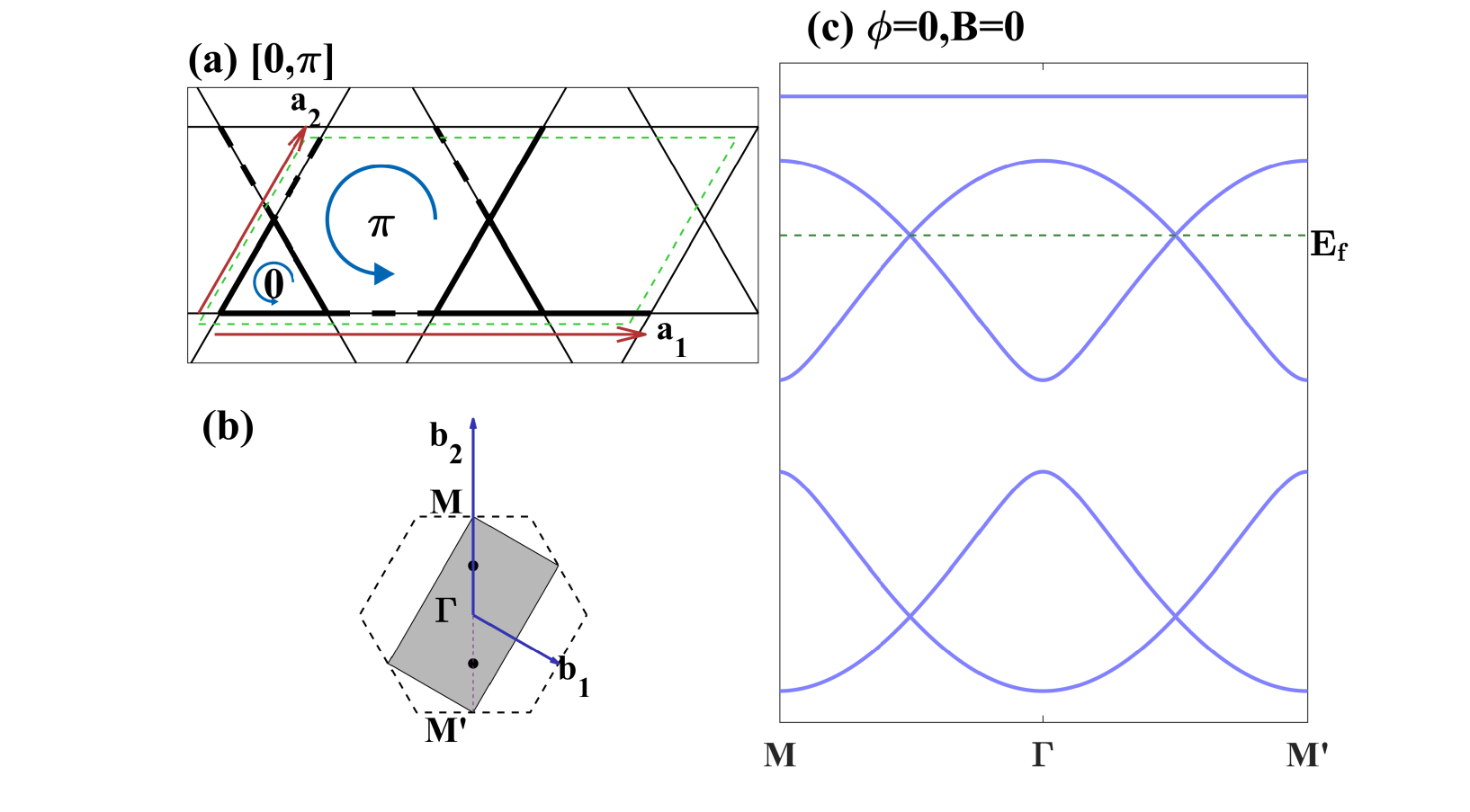}
 \caption{The U(1) gauge flux distribution and energy bands of $[0,\pi]$ DSL.
 (a) The flux pattern for $[0,\pi]$ DSL state. 
 $t_{ij}=\eta t$, ${\eta=\pm 1}$ for bold (dashed) lines. 
 $\bf{a_1,a_2}$ is the enlarged unit cell basis vectors.  
 (b) Brillouin zone (grey rectangle region) for the enlarged unit cell. 
 (c) Energy bands of $[0,\pi]$ state along $M-M'$ line for one spin sector.}
 \label{fig:2}
 \end{figure}

Combining variational Monte Carlo (VMC) and Gutzwiller projection~\cite{ran2007projectedwavefunction}, 
the U(1) DSL with the $[0,\pi]$ flux pattern shown in Fig.~\ref{fig:2}(a) gives the lowest variational 
energy of antiferromagnetic Heisenberg Hamiltonian, where the `0' flux refers to the flux 
through the corner triangular plaquette and the `$\pi$' flux refers to 
the flux through the central honeycomb plaquette. 
The spinon mean-field Hamiltonian of this U(1) DSL is written as 
\begin{equation}
\label{eq:Hmf}
\mathcal H_{\text{MF}}=-\sum_{\langle ij \rangle}{t_{ij} 
f^{\dagger}_{i\sigma}f_{j\sigma}} ,
\end{equation}
where the spinon hopping parameter $t_{ij}$ is specified in Fig.~\ref{fig:2}(a). 
With the half filling, the spinon Fermi level $E_f$ lies exactly at the Dirac cones 
in the band structure in Fig.~\ref{fig:2}(c). 
In the following, we take this $[0,\pi]$ U(1) DSL state as the parent state
and consider the effects of the flux generations. 
 
\section{Induced flux, spinon band reconstruction, and Berry curvature}
\label{sec:3}

\subsection{Staggered flux generation via the DM interaction}
\label{sec:3.1}

As we have mentioned in Sec.~\ref{sec:1} and Fig.~\ref{fig:DM}, 
the scalar spin chirality ${ \mathbf S_1\cdot (\mathbf S_2\times \mathbf S_3) }$, 
is related to the U(1) gauge flux~\cite{wen1989chiral,motrunich2006orbital}   
threading across the triangular plaquette formed by the sites 1,2,3, with 
\begin{equation}
\sin \Phi = \frac{1}{2}\mathbf S_1\cdot (\mathbf S_2\times \mathbf S_3).  
\label{eq:4}
\end{equation}
Often, this quantity cannot be directly tuned in experiments for the strong Mott insulators. 
For the weak Mott insulators, it is well-known from Motrunich's 
work~\cite{motrunich2005variational, motrunich2006orbital} that 
the strong charge fluctuation could generate a sizeable linear coupling 
between the scalar spin chirality and the magnetic field. 
This is referred to the orbital coupling effect of the magnetic field. 
In the strong Mott regime, the Zeeman coupling is the dominant coupling to the magnetic field. 
Remarkably, in a theoretical proposal by Lee and Nagaosa~\cite{lee2013proposal} 
that was aimed at probing the gauge field fluctuations via the neutron scattering, 
they observed that the presence of the DM interaction 
could generate the linear relation between the magnetic moments 
and the scalar spin chirality. 
Although their observation was about the experimental detection, 
it indirectly suggested the possibility of tuning the internal 
U(1) gauge flux via the simple external means like the Zeeman coupling. 
We here harness their observation and apply it to the $[0,\pi]$ U(1) DSL
for the kagom\'{e} lattice. 

\begin{figure}[t]
     \includegraphics[width=1\linewidth]{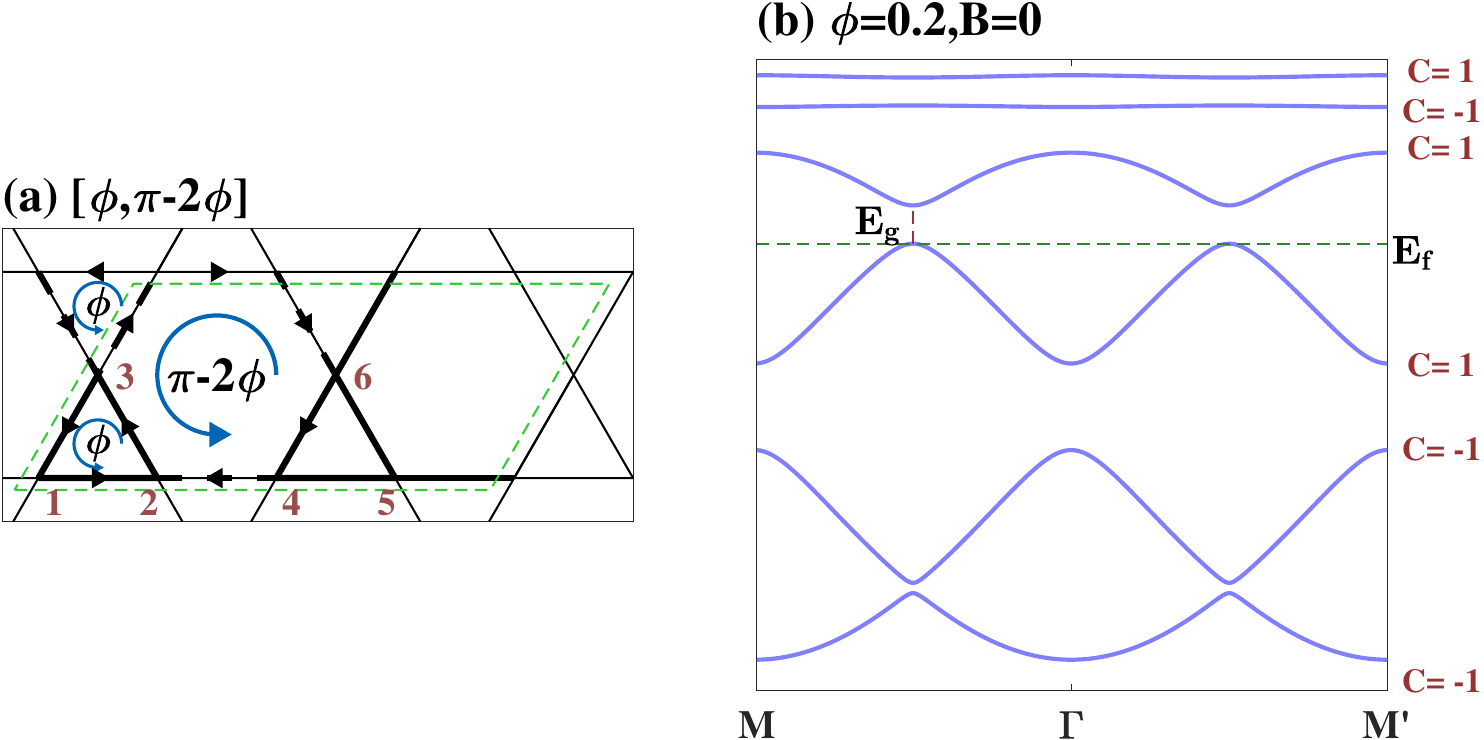}
 \caption{The U(1) gauge flux distribution and energy bands of one spin sector for the $[\phi,\pi-2\phi]$ flux pattern. (a) $[\phi,\pi-2\phi]$ flux pattern, where $t_{ij}=\eta_{ij} te^{i\zeta_{ij}\phi/3}$ and $\eta_{ij}=\pm 1$ for bold (dashed) lines. $\zeta_{ij}=1$ if the hopping from $j$ to $i$ is along the arrows and $\zeta_{ij}=-\zeta_{ji}$. (b) Gapped energy bands along the $M-M'$ line. When $\phi\neq 0$, the Dirac points near the Fermi surface will open a gap $E_g$ and all the bands become topological with a finite Chern number.}
 \label{fig:Ebphi}
 \end{figure}

Since the DM interaction appears in the Hamiltonian, 
a finite vector spin chirality ${\mathbf S}_i \times {\mathbf S}_j$ 
is induced for each bond, and this result is independent of the actual ground state. 
Once the spin on the neighbouring site that faces this bond is 
polarized by the external magnetic field, a finite scalar spin chirality could, 
in principle, be induced, though the reality depends on the field direction and 
the orientation of the DM vector. 
For the combination of the out-of-plane DM vector $D_z$ and 
the Zeeman coupling, a finite scalar spin chirality is induced 
for each triangular plaquette of the kagom\'{e} lattice. 
Along this line of reasoning, 
one has $\langle {\mathbf S}_2 \times {\mathbf S}_3 \rangle  \simeq \zeta \, {\mathbf D}_{23}$, 
where $\zeta$ is a proportionality constant and is roughly set by $1/J$. 
Moreover, the magnetization for the $z$-direction Zeeman field 
is ${M_z=\langle S^z_1 \rangle \simeq \chi B^z }$, 
where $\chi$ is the uniform magnetic susceptibility.

The uniform magnetic susceptibility requires a few more explanations here.  
Unlike the spinon Fermi surface state that has a constant magnetic susceptibility, 
the magnetic susceptibility vanishes for the U(1) DSL. 
In the presence of the generic DM interaction beyond the out-of-plane DM interaction, 
however, the global spin rotational symmetry is absent. 
Although the rotation around the $z$ axis remains to be a good symmetry for the model
with only the out-of-plane DM interaction,
the magnetization for the more general DM interaction 
is not a good quantum number to label the many-body states,
and the uniform magnetization is expected to be a small constant.

Combining the above results, we obtain the U(1) gauge flux through the triangular 
plaquette with the corner sites 1,2,3 as 
\begin{equation}
\label{eq:phiB}
\sin\phi \approx\frac{3}{2}\zeta D_z\langle S^z \rangle
=\frac{3}{2}\zeta D_zM_z   
\end{equation}
Since $|\zeta D_z|$ is small and the induced magnetization is also quite small, 
the induced U(1) gauge flux on each triangular plaquette is expected to be small.

At the level the parton mean-field theory, we further fix the gauge by assigning an extra $\phi/3$ phase
on each bond of the triangle anticlockwise (see Fig.~\ref{fig:Ebphi}a),
where $\phi$ is the induced gauge flux on the triangular plaquette.
Such a gauge fixing modifies the total flux on the central hexagon 
plaquette to be ${\pi-2\phi}$. The resulting flux pattern changes from $[0,\pi]$ to $[\phi, \pi-2\phi]$.
At this mean-field level, the net gauge flux on each unit cell of the kagom\'{e} lattice 
remains to be $\pi$, and stays the same as the one in the DSL~\cite{ran2007projectedwavefunction}.

\subsection{Spinon continuum for the magnetized $[\phi,\pi-2\phi]$ state}

Given the modified U(1) gauge flux pattern in Fig.~\ref{fig:Ebphi}, 
we proceed to explore the spinon mean-field model.  
The modified spinon-gauge-coupled Hamiltonian of $[\phi,\pi-2\phi]$ 
state is given as 
 \begin{equation}
 \label{eq:tightbind}
    \begin{aligned}
    \mathcal{H}_{\text{MF}}[\phi]=&-\sum_{\langle ij\rangle,\sigma}
                           (\eta_{ij}te^{i\gamma_{ij}\phi/3}f^{\dagger}_{i\sigma}f_{j\sigma}^{} +H.c.)\\
    &-\mu\sum_{i,\sigma} f^{\dagger}_{i\sigma}f^{}_{i\sigma}
    - B \sum_{i,\alpha\beta}f^{\dagger}_{i\alpha} \frac{\sigma^z_{\alpha\beta}}{2} f_{i\beta}^{} ,
    \end{aligned} 
\end{equation}
where ${\eta_{ij}=\eta_{ji}=\pm 1}$ and ${\gamma_{ij}=-\gamma_{ji}=\pm 1}$. 
Moreover, $\eta_{ij}= 1$ ($-1$) when the hopping is on the bold (dashed) bonds, 
and $\gamma_{ij}= 1$ when the hopping is along the arrow in Fig.~\ref{fig:Ebphi}(a). 
The chemical potential $\mu$ is introduced to constrain the particle number of the system.

\begin{figure}
    \centering
    \includegraphics[width=0.95\linewidth,trim=70 30 70 0,clip]{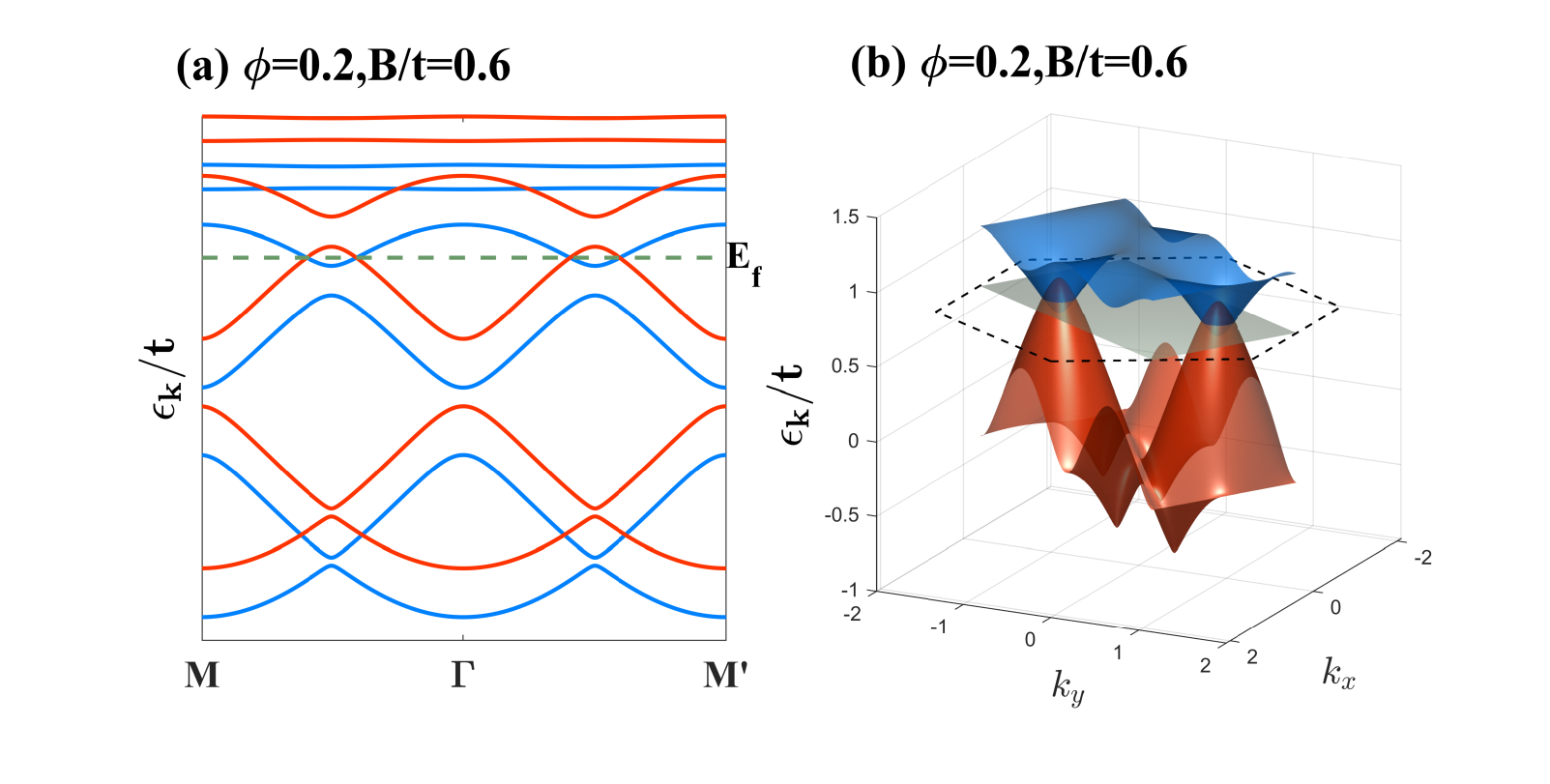}
    \caption{Spinon energy bands and Fermi pockets with finite staggered flux $\phi$ and a finite magnetization. 
    (a) Spinon energy bands along $M-M'$ line of the magnetized $[\phi,\pi-2\phi]$ state. 
    Blue (red) bands are the bands of the spin-up (down) sector.  
    (b) The nearest spin-up and spin-down bands near the Fermi surface cross each other, 
    forming Fermi pockets and resulting in finite magnetization.
    A relatively large $B$ field is chosen such that the band splitting and the outcome are more visible, and this applies
    to the remaining figures. }
    \label{fig:energybands}
\end{figure}

The induction of the staggered flux $\phi$ causes the Dirac cones 
near the Fermi surface to acquire a finite mass and open an energy gap $E_g$ 
(see Fig.~\ref{fig:energybands}(b)). 
Since the generation of $\phi$ requires finite magnetization (Sec.~\ref{sec:3.1}), 
the Zeeman energy shift needs to be included. 
With a sufficiently weak DM interaction, the Zeeman energy is expected 
to be large enough to shift the spin-up and spin-down spinon bands and 
create the Fermi pockets. This condition is given as
 \begin{equation}
 \label{eq:gap}
    |B/t|>E_g/t=2\left(\sqrt{3}+1\right)\sin(\frac{\phi}{3}),
 \end{equation}
 which is further justified in our numerics of Sec.~\ref{sec:5}. 
 
 \begin{figure}[b]
    \includegraphics[width=0.95\linewidth]{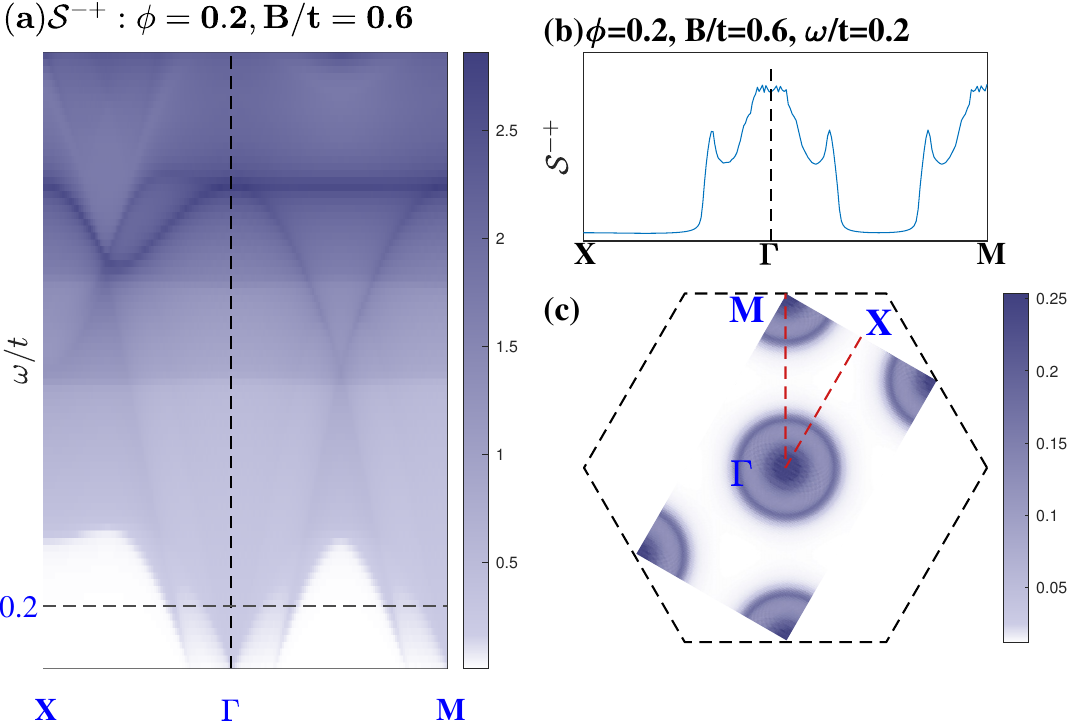}
    \caption{Spin dynamic structure factor $\mathcal S^{-+}$ of 
    massive FP state ( magnetized $[\phi,\pi-2\phi]$ state ). 
    (a) The continuum spectrum of $\mathcal S^{-+}$ along $X$-$\Gamma$-$M$ line.
    (b) $S^{-+}$ with constant energy $\omega/t=0.2$.
    (c) The momentum dependence of $S^{-+}( q, \omega)$ 
    at $\omega /t =0.2$ with $B/t=0.6$. 
    There are concentric ring signals at $\Gamma, M$ points.}
    \label{fig:INS}
\end{figure}
 
This field polarized state from ${\mathcal H}_{\text{MF}}(\phi)$ in 
Eq.~\eqref{eq:tightbind} is referred as the massive Fermi pocket (FP) state, whose mean-field wavefunction is given as
\begin{eqnarray}
|\Psi_{\text{FP}}\rangle = \prod_{\epsilon_{n\mathbf k,\sigma}<E_f}f^{\dagger}_{n\mathbf k\sigma} |0\rangle,
\end{eqnarray}
where $|0\rangle$ is the vacuum state for the spinons, $\epsilon_{n\mathbf k\sigma}$ is the energy of the excitation $f^\dagger_{n\mathbf k\sigma}\ket{0}$ of Eq.~\eqref{eq:tightbind}, $n,\mathbf k,\sigma$ is the index of energy band, wave vector and the spin sector. $E_f$ is the Fermi energy, which is set to remain spinon half-filling.
The signature of the massive Fermi pockets would manifest experimentally 
through the inelastic neutron scattering (INS), that reflects 
the spinon band characteristics and is proportional to 
the dynamic spin factor structure~\cite{shen2016evidence,li2017detecting, balents2020collective}:
\begin{equation}
    \begin{aligned} 
    \mathcal{S}^{a\bar a}(\mathbf{q}, \omega) & =\frac{1}{N} \sum_{i, j} e^{i \mathbf{q} \cdot\left(\mathbf{r}_{\mathbf{i}}-\mathbf{r}_{\mathbf{j}}\right)} \int e^{i \omega t}\left\langle S_i^{a}(t) \cdot S_j^{\bar a}(0)\right\rangle d t \\ 
    & =\sum_n \delta\left(\omega-\left(E_n(\mathbf{q})-E_0\right)\right)|\bra{n}S^{\bar a}_{\mathbf q}\ket{\Omega}|^2 ,
    \end{aligned}
\end{equation}
where the ground state $\ket{\Omega}$ with energy $E_0$ has fully filled 
spinons below the Fermi energy.  
$S^{a}(\mathbf q)\equiv 
\frac{1}{2}\sum_{\mathbf k}f^{\dagger}_{\mathbf{k-q},\alpha}\sigma^a_{\alpha\beta}f_{\mathbf{k},\beta} $, 
where ${a=\pm,z}$; ${\bar a=-a}$ for ${a=\pm}$, 
and $a=\bar a$ for $a=z$; $\sigma^+\equiv \sigma_x+i\sigma_y$.

\begin{figure}[b]
    \centering
    \includegraphics[width=0.96\linewidth]{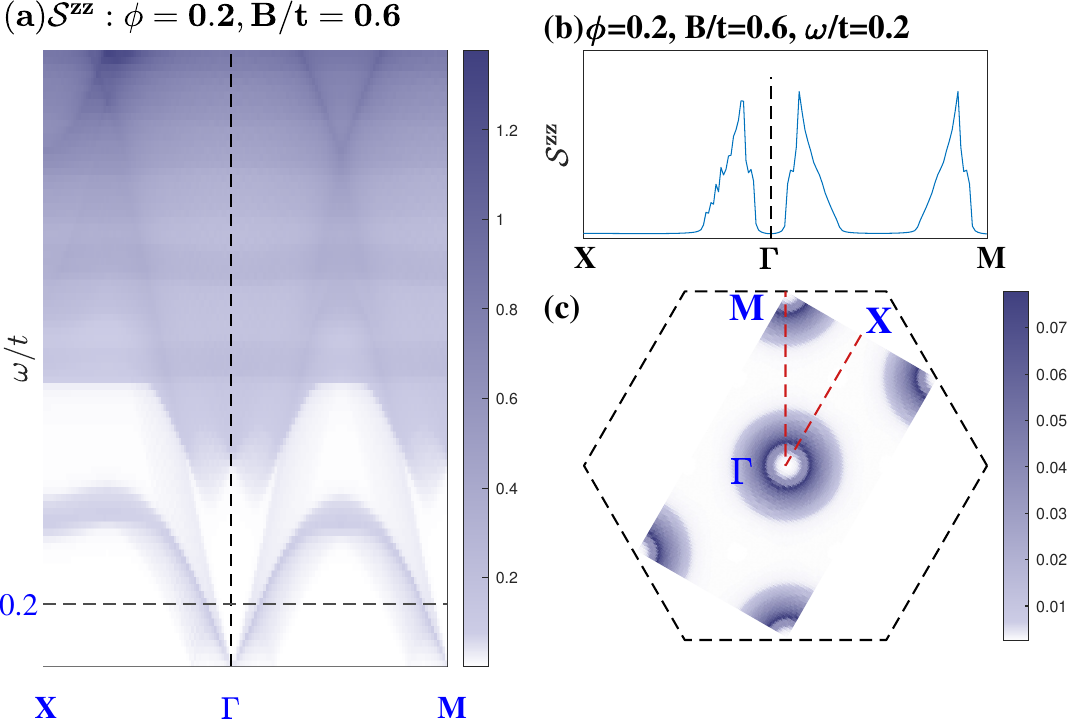}
    \caption{Spin dynamic structure factor $\mathcal S^{zz}$ of massive FP state 
    (magnetized $[\phi,\pi-2\phi]$ state). 
    (a) The continuum spectrum of $\mathcal S^{zz}$ along $X$-$\Gamma$-$M$ line.
    (b) $S^{zz}$ with constant energy $\omega/t=0.2$.
    (c) The momentum dependence of $S^{zz}( q, \omega)$ at $\omega /t =0.2$ with $B/t=0.6$.}
    \label{fig:INS_ZZ}
\end{figure}

As a practice, we calculate the dynamic spin structure factor and reveal the spinon
continuum for the FP state. Since the DM interaction and the Zeeman coupling 
break the global $SU(2)$ spin symmetry, $\mathcal S^{-+}$ and $\mathcal{S}^{zz}$ 
exhibit different behaviours. The first common feature is the gapless continuous 
spectrum, arising from the finite Fermi pockets. The second feature is about the 
spectrascopic signals at the $\Gamma$, $M$ points due to the intra-cone 
and inter-cone scattering across the Fermi surface.

The distinct features between $\mathcal S^{-+}$ and $\mathcal S^{zz}$ arise from
the different spinon scattering channels. 
$\mathcal S^{-+}$ reflects the spinon excitations across the Fermi surface between different spin sectors.
So a central peak signal surrounded by concentric rings emerges at both the $\Gamma$ and $M$ points illustrated in Fig.~\ref{fig:INS} because Fermi pockets will suppress hopping at certain momenta. For the real 
spin models, the spinon-interaction will further suppress the central peak signal at $\Gamma$ point.
For $\mathcal S^{zz}$, the spinon-hoppings between the same spin sector are allowed. Therefore, at low energies, peaks only appear on rings surrounding the $\Gamma$ and $M$ points.

\subsection{Thermal Hall effect of the $[\phi,\pi-2\phi]$ state}

The second effect of flux redistribution is to modify the Berry curvature, 
thereby creating the thermal Hall effect~\cite{qin2011energy,zhang2024thermal,gao2020topological}.
For a non-interacting fermion system, the thermal Hall conductivity is given as
\begin{equation}
\label{eq:kappa}
    \kappa_{x y}=-\frac{1}{T} \int d \epsilon(\epsilon-\mu)^2 
    \frac{\partial f(\epsilon, \mu, T)}{\partial \epsilon} \sigma_{x y}(\epsilon),
\end{equation}
where ${f(\epsilon, \mu, T)=1 /\left[e^{\beta(\epsilon-\mu)}+1\right]}$ is the Fermi-Dirac distribution function, 
and $\sigma_{x y}(\epsilon)=-1 / \hbar \sum_{\mathbf{k}, \sigma, \epsilon_{n, \mathbf{k}}<\epsilon} \Omega_{n, \mathbf{k}, \sigma}$. Here, $\Omega_{n,\mathbf{k}}$ is the Berry curvature of the single-particle excitation $\ket{n,\mathbf k}$. 
When the temperature is approaching zero limit, 
$\kappa_{xy}/T$ of the gapped system is integer quantized in units of $-\frac{\pi k_B^2 }{6\hbar}$: 
\begin{equation}
\left.\dfrac{\kappa_{xy}}{T}\right|_{T\rightarrow 0}=-\dfrac{\pi k_B^2 }{6\hbar}\sum_{n\in\text{filled bands}}C_n ,
\end{equation}
where $C_n$ is the Chern number of the $n$-th band~\cite{zhang2024thermal}.

\begin{figure}[t]
    \centering
    \includegraphics[width=0.92\linewidth]{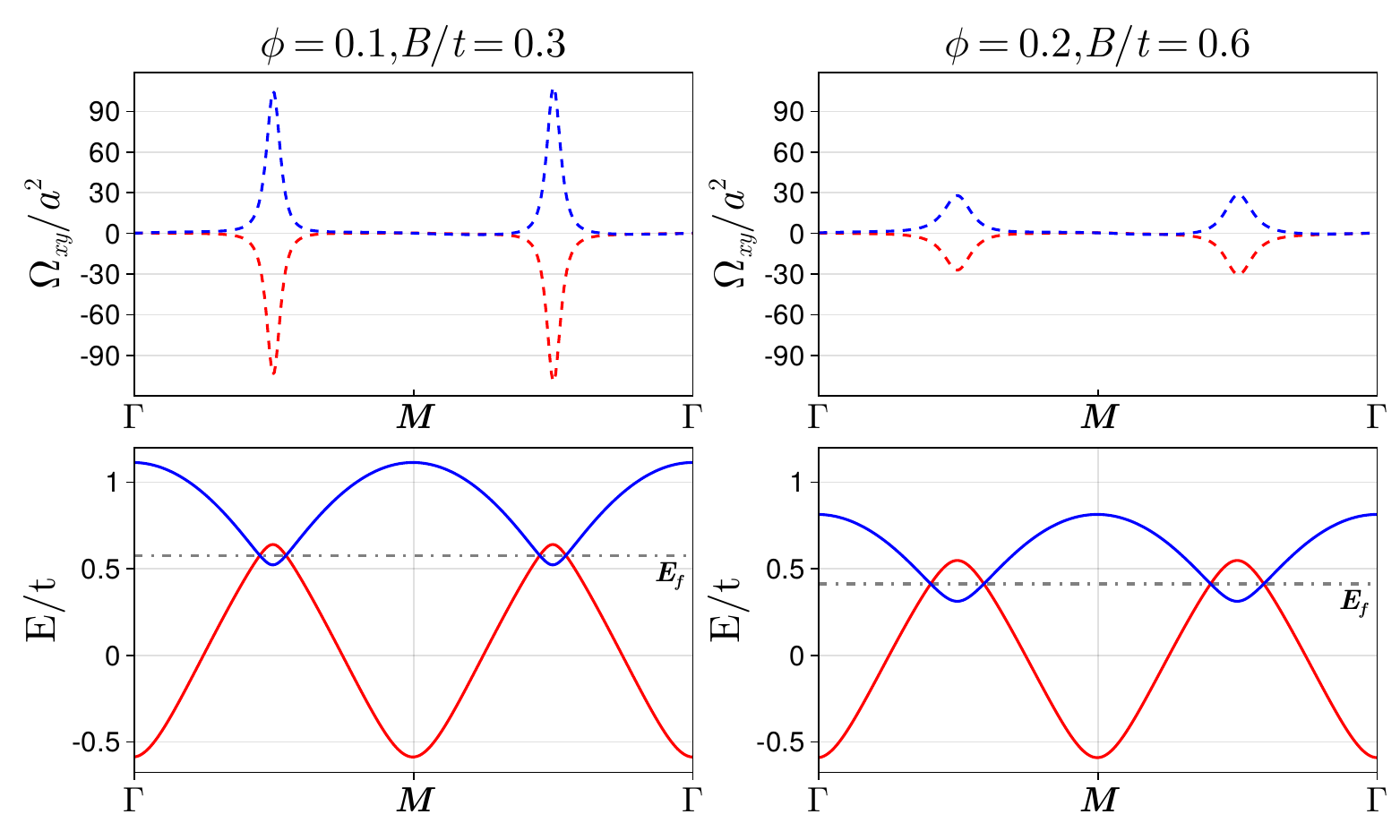}
    \caption{Berry curvature of the massive Dirac cones near the Fermi surface.
    The top panel figures are the Berry curvatures of the two bands nearest to Fermi surface along $\Gamma-\Gamma'$ under different flux $\phi$, with the magnetic field $B/t=3\phi$. The bottom panel figures are the corresponding energy bands. Blue (red) bands are the band of the spin-up (down) sector. Here $a$ denotes the nearest-neighbor lattice constant.}
    \label{fig:bcline}
\end{figure}

In the presence of a finite staggered field $\phi$ but vanishing Zeeman coupling, 
the spin-$\uparrow$ and spin-$\downarrow$ spinon bands in the mean-field treatment are degenerate,
and all bands have gaps from each other. The Chern numbers of these spinon 
bands in Fig.~\ref{fig:Ebphi}(b) are $\{-1,-1,1,1,-1,1\}\times \text{sign}(\phi)$ for both spin sectors.
This band scheme would contribute to a non-trivial Hall effect with quantization 
at zero-temperature limit at half-filling ~\cite{maity2025thermal}. 
With the Zeeman coupling, however, the Fermi pockets form 
and show massive Dirac cones with opposing Berry curvatures near the Fermi surface in Fig.~\ref{fig:bcline}. 
Therefore, the spin-$\uparrow$ Fermi pocket below the Fermi surface not only fails to 
compensate for the Berry curvature contribution from the spin-$\downarrow$ hole pocket, 
but also reduces the total thermal Hall conductivity, 
leading to non-quantized behaviour~\cite{teng2020unquantized}.

\begin{figure}[b]
    \centering
    \includegraphics[width=1.0\linewidth]{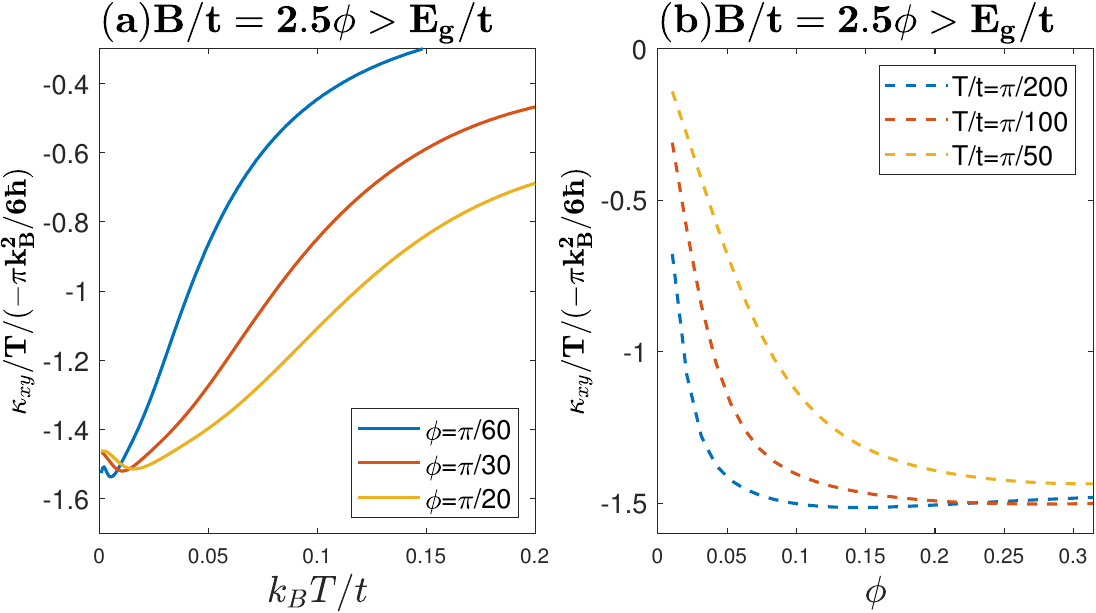}
    \caption{(a) Temperature dependence of $\kappa_{xy}/T$ for various values of $\phi$. 
    (b) Flux $\phi$ dependence (magnetic field dependence) of $\kappa_{xy}/T$ for various temperatures.}
    \label{fig:kappa_line}
\end{figure}

This spinon Fermi pocket is reflected in 
the temperature dependence of $\frac{\kappa_{xy}}{T}/(-\frac{\pi k_B^2}{6\hbar})$, 
showing an initial decrease followed by an increase that asymptotically approaches zero. 
This is because as $T$ increases, the rising chemical potential first incorporates 
the negative $\Omega(\mathbf k)$ of the spin-$\downarrow$ hole Fermi pocket into the total $\kappa_{xy}$, 
leading to the initial decrease in the thermal Hall conductivity.
 Subsequently, with further increase of $\mu$, contributions from all spinon bands 
 drive $\kappa_{xy}/T$ asymptotically toward zero (Fig.~\ref{fig:kappa_line}(b)).

Another characteristic of the Fermi pocket is shown in the dependence of $\kappa_{xy}/T$ on $\phi$. 
When $\phi$ approaches zero, the mass of the Dirac cones tends to zero, 
causing $\Omega(\mathbf k)$ of the spinon bands near the Fermi surface 
to concentrate within the Fermi pockets. 
After the positive Berry curvature of the spin-$\uparrow$ Fermi particle pockets 
cancels the contribution of other bands below the Fermi surface, 
the total $\kappa_{xy}/T$ will tend to zero. 
So if we switch on the staggered flux $\phi$, 
the Berry curvature will propagate outward from the Dirac points, 
and then $\left|\kappa_{xy}/T\right|$ will gain a finite value. 
If the flux $\phi$ is large enough, the Zeeman shift causes the Fermi pockets 
to expand at a rate faster than the Berry curvature, 
leading to a decrease in $\left|\kappa_{xy}/T\right|$ (Fig.\ref{fig:kappa_line}(b)).

\section{Spontaneous flux generation and massive Landau level state}
\label{sec:4}

The previous section discussed how the U(1) gauge flux is perturbatively modified 
by the DM interaction and the weak magnetization. Our analysis addresses 
the gauge field configuration with a net flux of $\pi$ per unit cell 
of the kagom\'{e} lattice, and the emergent U(1) gauge field remains gapless.
Nevertheless, the finite spinon Fermi surface of the massive FP state
may lead to other instabilities. 
It is likely that the system spontaneously generates a finite U(1) gauge flux 
and gains the kinetic energy for the spinon matter by creating and 
filling the spinon Landau levels. This possibility has been
considered by Y. Ran et al for the massless Dirac spinons in the kagom\'{e}
U(1) DSL in the weak magnetic field~\cite{ran2009spontaneous}. 
Here, we consider such a possibility for the FP state where the Dirac cones 
have a finite mass gap.

\begin{figure}[t]
    \centering
    \includegraphics[width=0.95\linewidth]{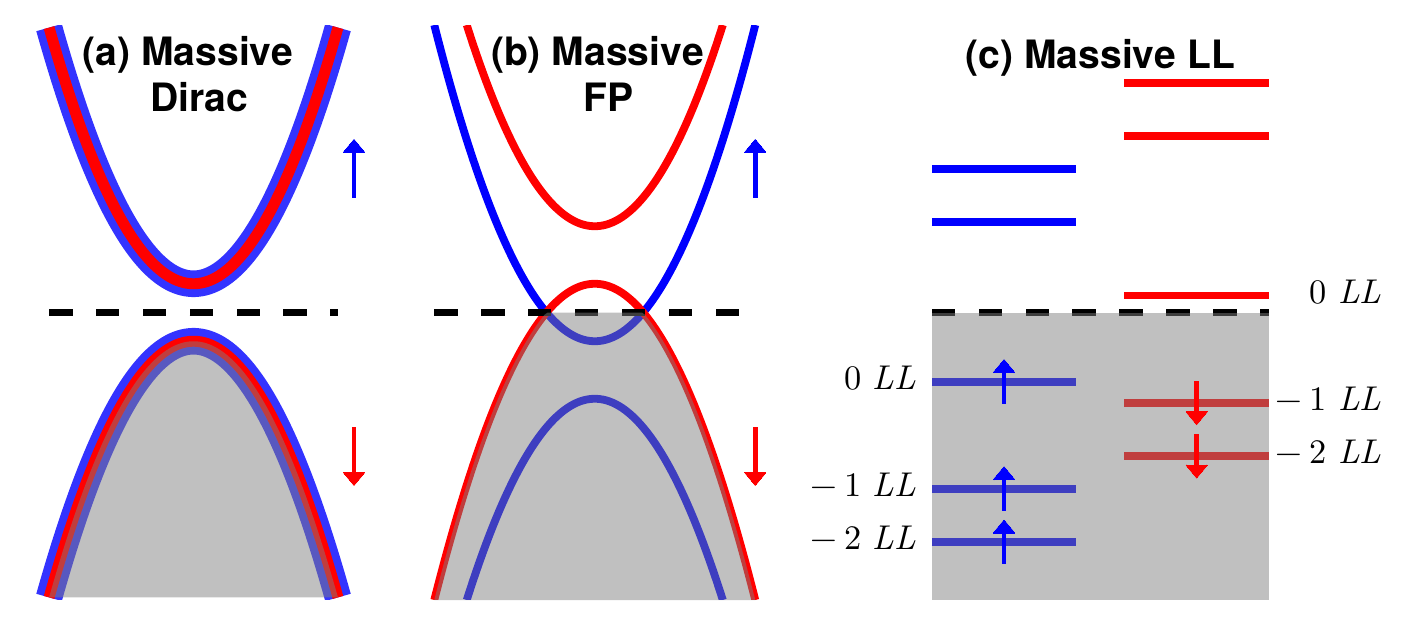}
    \caption{The Fermi sea of the massive Dirac cone state (massive Dirac) in (a), 
    the massive Fermi pocket (FP) state in (b) and the massive Landau Level (LL) state in (c). 
    The black dashed line (gray shaded area) represents the Fermi surface (filled fermion states). 
    Blue (red) lines represent spin-up (spin-down) spinon excitations. 
    The internal U(1) flux $\mathbf b$ is chosen such that $\varepsilon_0<0$ in Eq.~\eqref{eq:epsilon_spec}. 
    Under the Zeeman effect, $0$-LL is fully occupied by spin-up spinons and fully empty by spin-down spinons, 
    such that the massive FP and the massive LL states have the identical magnetization $M_z$. }
    \label{fig:LL_filling}
\end{figure}

To define this spinon Landau level state from the spontaneous flux generation, 
we start from the low-energy effective theory of the massive Dirac fermions 
coupled to a U(1) gauge field ${\mathbf a}$ in $2+1$D for the FP state 
as: 
\begin{equation}
	\mathcal{H}_{md}^{\pm}(\mathbf{q-a},m)=v_{f}^{\pm}
	\left[(q_{x}-a_x)\sigma_{x}+(q_{y}-a_y)\sigma_{y}\right] 
	+m^{\pm}\sigma_{z},
\end{equation}  
where $\pm$ is the index of the two Dirac cones near the Fermi surface 
of the $[\phi,\pi-2\phi]$ state,  
$v_f^\pm$ is the Fermi velocity, and $m^\pm$ is the mass of the Dirac fermions. 
The Zeeman splitting of the spin-$\uparrow$ and the spin-$\downarrow$ spinons has not yet
been included. 
For the two nodes, they share the same chirality, 
which means that $v_f^\pm=v_f$ and $m^\pm=m=-(1+\sqrt{3})t\sin({\phi}/{3})$. 
Therefore, the $\pm$ node index will be omitted in the following discussion.

With a uniform spontaneous U(1) gauge flux, one 
fixes the gauge field vector as $\mathbf a=(0,-xb)$ for the internal field 
$\mathbf b=-b\mathbf{e}_z$, and obtains the spectrum of the Landau levels:
\begin{eqnarray}
\label{eq:epsilon_spec}
&&	  \varepsilon_0=-\text{sgn}(b)\sqrt{2|b|}v_{f}\Delta=-\text{sgn}(bm)\times |m| ,    
\\
	&& \varepsilon_n=\text{sgn}(n)\sqrt{2|b|}v_{f}\sqrt{\Delta^{2}+|n|},
     \quad n=\pm 1,...,
\end{eqnarray}
with $\Delta={m}/{(\sqrt{2|b|}v_{f})}$. Without the loss of generality,  
one can always set $\varepsilon_{0}$ of the two nodes to be negative 
at the same time with a specific direction of $\mathbf{b}$.

In the presence of Zeeman coupling, the choice of the spontaneous flux is that 
the system fills the spin-$\uparrow$ spinons up to the $0$-th Landau level (0-LL) 	
and fills the spin-$\downarrow$ spinons up to the $-1$-LL in Fig.~\ref{fig:LL_filling}.
Since the state arises from the massive Dirac cones, we dub the 
 resulting state the massive Landau level (LL) state simply to distinguish it from the massless
 LL state out of the massless Dirac cone, and the mean-field wavefunction 
 for this state is 
given as 
\begin{equation}
|\Psi_{\text{LL}}\rangle \equiv  \prod_{\eta=\pm}\prod_{n\leq 0}\prod_{m\leq -1}\prod_{l}f^{\eta\dagger}_{nl\uparrow}f^{\eta\dagger}_{ml\downarrow}   \ket{0}. 
\end{equation}
Here, $n,m$ are the integer indices of the Landau levels. $\eta=\pm$ is the index of the two Dirac cone nodes near the Fermi surface of U(1) DSL. $l$ is the index of the degeneracy of the Landau levels.
 To match the particle number with the energy levels, 
 we can infer that the sign and magnitude of 
 internal field $\mathbf b$,
\begin{equation}
\label{eq:b}
    \left\{
    \begin{aligned}
       &\text{sign}(b)=-\text{sign}(\phi),\\
       &\Delta n=\frac{N_\uparrow-N_\downarrow}{A}=n_{\uparrow}-n_{\downarrow}=2\frac{b}{2\pi}=2D/A,
    \end{aligned}
    \right.
\end{equation}
where $n_{\uparrow}, n_\downarrow$ are the densities of the spin-$\uparrow$ 
and spin-$\downarrow$ spinons, $D/A$ is the density of states of each Landau level of one node.

Here we further compare the mean-field energies of the FP and LL states with the Zeeman coupling. 
We set the gapless DSL as the reference state and compare the energy changes of the system 
for the LL and the FP states. We have 
\begin{eqnarray}
&& \delta\varepsilon_{\text{LL}} = \varepsilon_{\text{LL}} - \varepsilon_{\text{DSL}} + B m^z ,\\
&& \delta\varepsilon_{\text{FP}} = \varepsilon_{\text{FP}} - \varepsilon_{\text{DSL}} + B m^z ,
\end{eqnarray}
where $\varepsilon_{\text{LL}}$, $\varepsilon_{\text{FP}}$ and
$\varepsilon_{\text{DSL}}$ are the mean-field energies of the corresponding states per
site, and $m^z$ is the magnetization per site with $ B m^z $ taking care of the Zeeman
energy. After some calculation, we find 
that 
\begin{equation}
\label{eq:ratio}
    \frac{\delta \varepsilon_{LL}}{\delta\varepsilon_{FP}}=\frac{3\sqrt{2} 
     \left[-\frac{2|\Delta|^3}{3}-\zeta(-\frac{1}{2},1+\Delta^{2})-\frac{|\Delta|}{2}\right]}{  (1+2\Delta
    ^{2})^{\frac{3}{2}}-(2\Delta^{2})^{\frac{3}{2}} } ,
\end{equation}
where $\zeta(-\frac{1}{2},1+\Delta^{2})$ is the zeta function.
The ratio $\delta\varepsilon_{FP}/\delta \varepsilon_{LL}$ decreases 
monotonically with increasing $|\Delta|$, which is always smaller than 1. 
Thus, the LL state is the ground state of the system with a spontaneous 
internal flux, instead of the FP state.

At this stage, we have shown within the mean-field theory that,  
the massive LL state induced from the spontaneous flux generation 
on top of the perturbative flux from the DM interaction and Zeeman coupling
exhibits a lower mean-field energy than the massive FP state. 
The presence of the Dirac mass gap from the combined effect of the DM interaction 
and the magnetization actually makes the system energetically more susceptible 
to the spontaneous flux generation for the emergence of the massive LL state. 
Finally, the spontaneously generated internal U(1) gauge flux is related to the 
direction and strength of the out-of-plane magnetization,
as described by Eqs.~\eqref{eq:phiB} and \eqref{eq:b}.

\begin{figure}[t]
    \centering
    \includegraphics[width=0.7\linewidth]{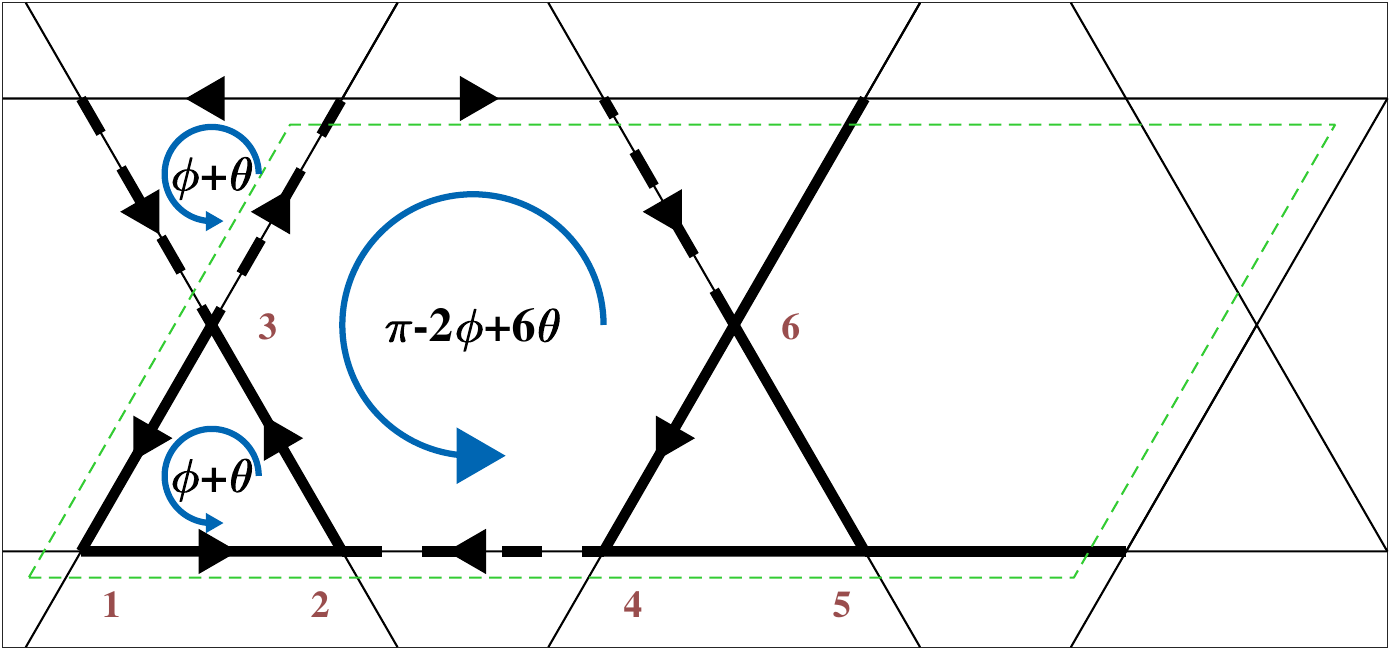}
    \caption{Flux pattern for [$\phi+\theta$,$\pi-2\phi+6\theta$] state. 
    $\phi$ is the staggered flux, $\theta$ is the background uniform flux in each triangle plaquette, 
    with $6\theta$ through each hexagonal plaquette. 
    The corresponding hopping parameter $t_{ij}=\eta_{ij} te^{i\zeta_{ij}\phi/3}e^{i\theta_{ij}}$ 
    and $\eta_{ij}=\pm 1$ for bold (dashed) lines. 
    $\zeta_{ij}=1$ if the hopping from $j$ to $i$ is along the arrows and $\zeta_{ij}=-\zeta_{ji}$. 
    $\theta_{ij}$ is used to construct the background uniform flux configuration of $\theta$, 
    with its selection referring to Ref.~\onlinecite{du2018floquet}.  
    }
    \label{fig:LL_flux}
\end{figure}

\begin{figure*}[th]
    \centering
    \includegraphics[width=0.85\textwidth]{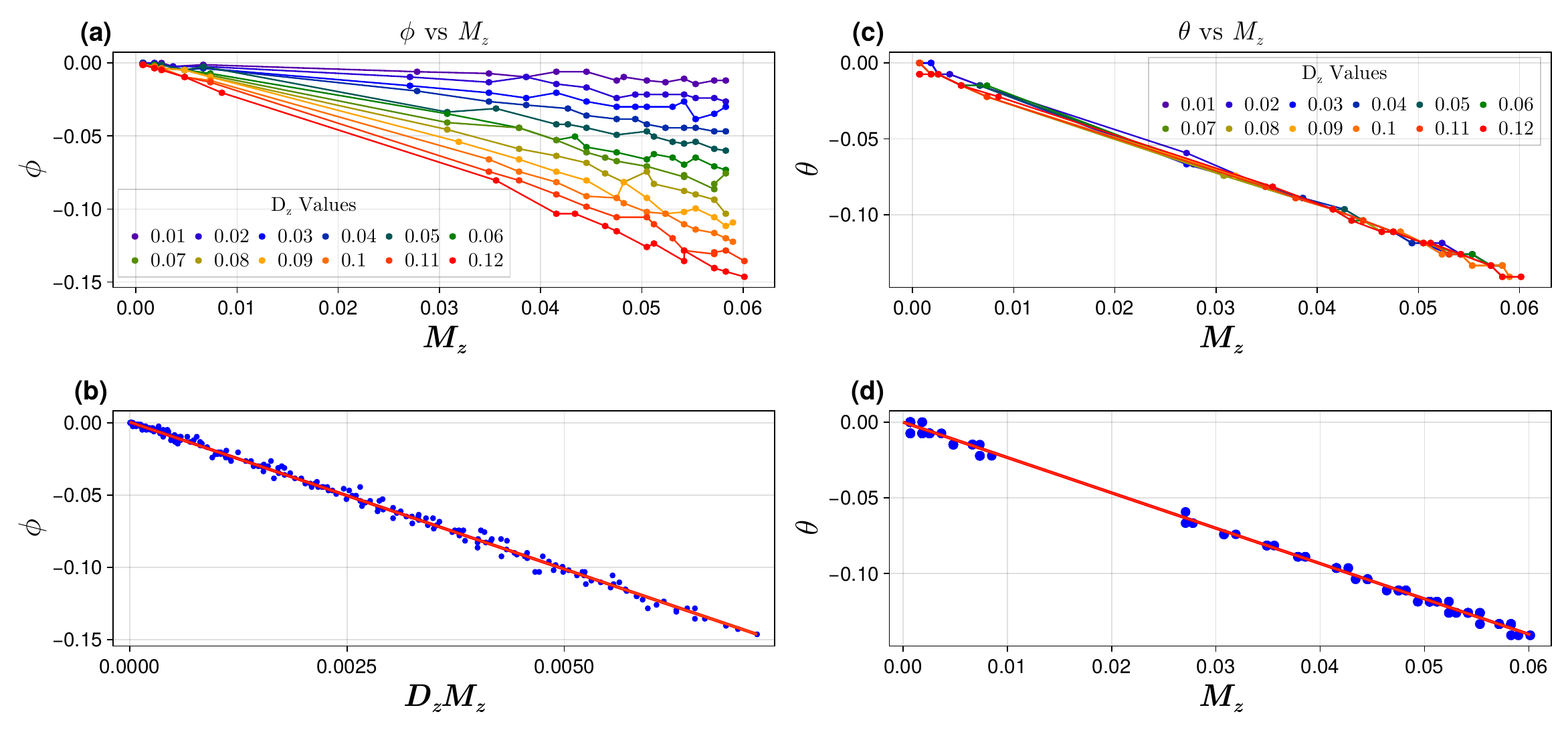}
    \caption{RMFT optimization results of the massive FP state and the massive LL state.
    The massive FP state  ($[\phi,\pi-2\phi]$ state) was optimized by varying $\phi$ and $M_z$
     across different $D_z$ and $B$ to the lowest energy $E_z$. 
     (a) shows the resulting $\phi$ versus $M_z$ dependence.
    The linear fit ${\phi=\eta_\phi D_zM_z+\beta_\phi}$ in (b) yields the slope $\eta_\phi=-20.36(7)$ 
    with a negligible intercept $\beta_\phi=0.0006(2)$, where blue dots and the red line denote 
    the RMFT raw data and the linear fit, respectively.
    Treating this massive FP state as the parent state, 
    $\theta$ is modified to gain the lowest energy of $[\phi+\theta,\pi-2\phi+6\theta]$ state 
    with the given $\phi$ and $M_z$ of the corresponding $D_z$ and $B$.  
    (c) shows the $M_z$-dependence of $\theta$ under different $D_z$'s, revealing a universal scaling behaviour.
    The linear fit in (d) yields ${\theta=\eta_{\theta}M_z+\beta_\theta}$ with $\eta_\theta=-2.3351(55)$ , 
    demonstrating an exclusive scaling with $M_z$ and no dependence on $D_z$ magnitude.  
    The negligible intercept $\beta_\theta=0.0000(2)$ further confirms this $M_z$-linear relationship.}
    \label{fig:RMFT}
\end{figure*}

\section{Verification of the Renormalized mean-field theory}
\label{sec:5}

\subsection{Renormalized mean-field theory}
\label{sec:5a}

In the previous sections, we have discussed how the combination of the out-of-plane 
DM interaction $D_z$ and the out-of-plane magnetization $M_z$ introduce a staggered flux, 
resulting in a Fermi pocket state of the KHAFM. 
Furthermore, we showed that the massive Landau level state induced 
by a spontaneous uniform flux is more stable in energy. 
Both aspects were studied based on the free-spinon mean-field theory.

Due to the Hilbert space constraint, we here perform the Gutzwiller 
projection~\cite{gros1989physics} for the mean-field ground state. 
The variational Monte Carlo method~\cite{he2024spinon,iqbal2016spin,
iqbal2021gutzwillerprojected, motrunich2005variational,ran2007projectedwavefunction,sorella2005wave} 
is typically employed to optimize the projected energy and to 
obtain the suitable mean-field parameters. To simplify the process, 
we apply the renormalized mean-field theory (RMFT)~\cite{gros1989physics, 
ogata2003superconductivity, zhang1988renormalised} 
to estimate the mean-field effects of the projection operator $\hat P_G$. 
The idea of the RMFT is to use a renormalization factor, 
that depends on the mean-field ground state $\ket{\psi_0}$,
to estimate the expectation of the operator $\hat O$ 
in the mean-field state $\ket{\psi_0}$ versus its expectation 
in the projected state $\ket{\psi}=\hat P_G\ket{\psi_0}$:
\begin{equation}
\label{eq:Gutz_factor}
    \frac{\bra{\psi_0} \hat P_G\hat O\hat P_G\ket{\psi_0}}{\bra{\psi_0} 
    \hat P_G\ket{\psi_0}}
    \approx g_O \bra{\psi_0} \hat O\ket{\psi_0},
\end{equation}
where $g_O$ is the Gutzwiller renormalization factor for $\hat O$.

In this method, one can evaluate the energy of the projected state 
$\hat P_G\ket{\psi_0}$ from a mean-field perspective. 
We write down the Hamiltonian with the out-of-plane DM interaction 
and the Zeeman effect as 
\begin{equation}
\begin{aligned}
\mathcal H_z&=-\sum_{i}BS^{z}_{i}+J\sum_{\braket{ij}}S^{z}_{i}S^{z}_{j}\\
&+\sum_{\braket{ij}}\left[\frac{J+iD^z_{ij}}{2}S^{+}_{i}S^{-}_{j}+\frac{J-iD^z_{ij}}{2}S^{-}_{i}S^{+}_{j}\right].
\end{aligned}
\label{eq21}
\end{equation}
Thus the energy of this system is calculated within the RMFT as
\begin{equation}
\label{eq:ProjEnergy}
    \begin{aligned}
        E_z
    &=\frac{\bra{\psi_0}\hat P_G\mathcal  H_z\hat P_G\ket{\psi_0}}{\bra{\psi_0} \hat P_G\ket{\psi_{0}}}\\
    &\approx-\sum_{i}B\braket{S^z_i}_0+g_{zz}J\sum_{\braket{ij}}\braket{S^z_iS^z_j}_0\\
    & \quad +\sum_{\braket{ij}} \Big(g_{+-}\frac{J+iD_z}{2}\braket{S^+_iS^-_j}_0+h.c.\Big),
    \end{aligned}
\end{equation}
where $\braket{\cdots}_0$ is the expectation value with respect to the mean-field ground state $\ket{\psi_0}$, and
$g_{zz}$, $g_{+-}$ are the Gutzwiller renormalization factors for $S^z_iS^z_j$ and $S^+_iS^-_j$, respectively.

In the RMFT, it is crucial to determine the Gutzwiller renormalization factors 
$g_{zz}$ and $g_{+-}$ for accounting various properties of the system, 
such as the antiferromagnetic order \cite{gros1989physics, ogata2003superconductivity, zhang1988renormalised}.
It has been shown that the zeroth-order approximation of the Gutzwiller factor, 
which only considers the on-site correction of the projection operator, 
is insufficient to capture the antiferromagnetic correlations, 
leading to identical $g_{zz}$ and $g_{\pm}$ \cite{ogata2003superconductivity}. 
The identical $g_{zz}$ and $g_{\pm}$ merely leads to an overall rescaling effect 
of the external magnetic field on top of the free-spinon mean-field results. 
Moreover, the identical $g_{zz}$ and $g_{\pm}$ is inconsistent with 
the U(1) symmetry of our spin model. 
Therefore, the inter-site correlation of the projection operator needs be included 
for a better description of the system.
As detailed in Appendix.~\ref{App:RMFT}, 
we evaluate the explicit forms of $g_{zz}$ and $g_{+-}$ 
by considering the real-space correlation upto three neighboring sites 
from the triangular plaquettes of the kagom\'{e} lattice to capture the 
effect of the underlying U(1) gauge flux. 
We obtain the expected different values for $g_{zz}$ and $g_{+-}$, 
which are functions of the mean-field parameters 
$\chi^\sigma_{ij}\equiv\braket{f^\dagger_{i\sigma}f_{j\sigma}}_0$ and $M_z$, 
and we discuss our RMFT results in the following subsection.

\subsection{RMFT results}
\label{sec:5b}

The $[\phi+\theta,\pi-2\phi+6\theta]$ state is used to give the mean-field ground state $\ket{\psi_0}$. 
The background uniform flux $\theta\equiv-bA_{\text{tri}}$ is added into each triangle plaquette 
(see Fig.~\ref{fig:LL_flux}), where $A_{\text{tri}}$ is the area of a triangular plaquette.
To satisfy the periodic boundary condition, we choose ${\theta={2\pi q}/(16p)}$ 
and an enlarged magnetic unit cell that is $p$ times larger than the original 
crystal unit cell along the $\mathbf a_1$ direction in Fig.~\ref{fig:2}. 
By modifying the parameters $\phi,\theta$ and the out-of-plane magnetization $M_z$, 
$\ket{\psi_0}$ can be optimized to achieve the lowest energy $E_z$ in 
Eq.~\eqref{eq:ProjEnergy} for different $D_z$ and magnetic fields $B_z$.

We first consider the energy optimization of the [$\phi$,$\pi-2\phi$] 
state without the uniform flux $\theta$, which explores the relationship 
between $\phi$ and $D_zM_z$ in the massive FP state. Since $\phi$ is 
expected to be proportional to $D_zM_z$ in the small $\phi$ limit, 
Fig.~\ref{fig:RMFT}(a)(b) shows that $\phi$ indeed exhibits a linear dependence on 
$D_zM_z$, with the fitting results as, 
\begin{equation}
\label{eq:LF_phi}
    \phi\approx -20.36(7)D_z M_z+0.0006(2).
\end{equation}
This relation agrees with the staggered flux in Eq.~\eqref{eq:phiB}.

With this result, we continue to employ the RMFT to explore the uniform U(1) gauge flux $\theta$.  
Specifically, for given $D_z$ and $B$ values, based on the optimized $\phi$ and the magnetization 
$M_z$ obtained from the aforementioned optimization of ${[\phi, \pi-2\phi]}$, 
we adjust the uniform gauge field $\theta$ to minimize the energy $E_z$. 
The first conclusion obtained from Fig.~\ref{fig:RMFT}(c) is that, 
the sign of $\theta$ is the same with the sign of $\phi$:
\begin{equation}
\label{eq:sign_theta_phi}
    \text{sign}(\theta)=\text{sign}(\phi),
\end{equation}
which agrees with the conclusion of Eq.~\eqref{eq:b}. 
Moreover, the strength of $\theta$ is only proportional to $M_z$ shown in Fig.~\ref{fig:RMFT}(d),
\begin{equation}\label{eq:LF_theta}
    \theta\approx -\text{sign}(D_z)2.3351(15)M_z+0.0000(2).
\end{equation}
This is consistent with the conclusion of Eq.~\eqref{eq:b}, which states that 
the magnetization $M_z$ determines the strength of the spontaneous uniform gauge field.
Thus, the RMFT confirms the mean-field results.

\begin{figure}[t]
    \centering
    \includegraphics[width=0.8\linewidth,trim=10 10 10 0,clip]{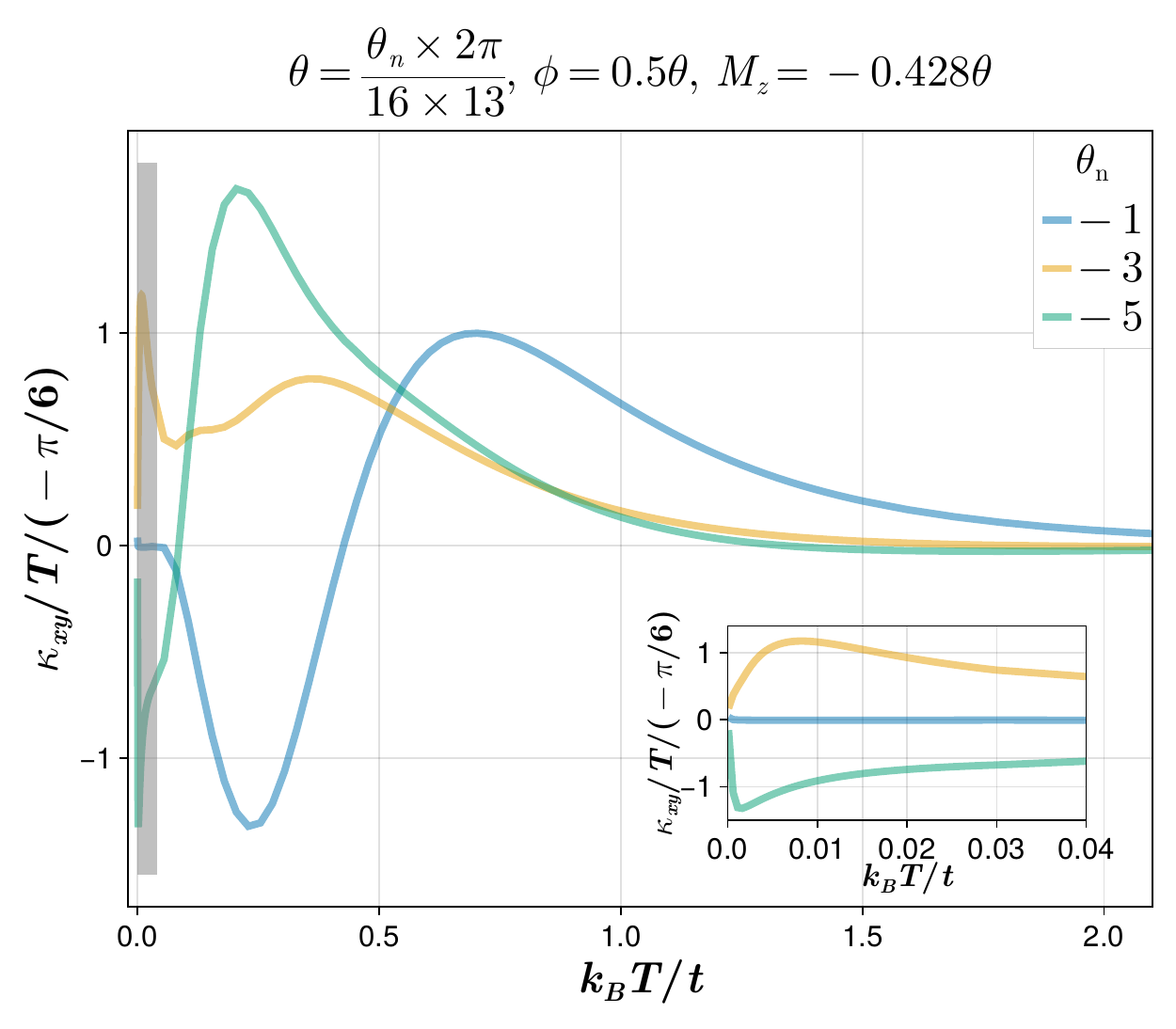}
    \caption{Temperature $T$ dependence 
    of the thermal Hall conductivity $\kappa_{xy}$ for the massive LL 
    state at fixed magnetization for given $\theta,\phi$.  
    The shaded area is magnified and plotted in the inset, 
    which depicts the variation of $\kappa_{xy}/T$ at low temperatures.}
    \label{fig:kappa_LL}
\end{figure}

Based on the renormalized mean-field optimization in Eq.~\eqref{eq:LF_theta}, 
we perform an analysis of the outcome in the massive LL state. 
With the uniform flux $\theta$, the original translational symmetry is 
replaced by a magnetic translational symmetry characterized by an enlarged unit cell 
in the spinon mean-field theory. Consequently, the spin dynamic structure factor 
will reflect the spinon Landau levels at the relevant energy scales.  
The spinon Landau bands acquire a non-trivial distribution of Berry curvatures. This 
feature manifests in the chemical-potential-sensitive thermal Hall effect.
Using the linear fitting results from Eqs.~\eqref{eq:LF_phi} and \eqref{eq:LF_theta}, 
we select and fix the relationship among $\theta$, $\phi$, and the out-of-plane magnetization $M_z$. 
As temperature rises, the chemical potential of both spin-up and spin-down spinons will 
increase and cross more Landau subbands compared to the FP state, 
leading to additional monotonic variations in $\kappa_{xy}/T$ that eventually 
approaches zero (Fig.~\ref{fig:kappa_LL}).

\section{Discussion}
\label{sec:6}

In this paper, we have shown that, the out-of-plane DM interaction   
not only induces the U(1) gauge flux but also stabilizes the massive 
Landau level state from the spontaneous flux generation by the external magnetic field 
for the U(1) DSL on the kagom\'{e} lattice. 
In the presence of the out-of-plane DM interaction and the out-of-plane 
magnetic field, the global U(1) symmetry is preserved in the spin model of Eq.~\eqref{eq21}.
Here we closely follow the argument of Ref.~\onlinecite{ran2009spontaneous}
and conclude the massive Landau level state is an ordered antiferromagnet 
with the in-plane antiferromagnetic order.
We then provide a bit more microscopic understanding and explain the excitations
of this ordered antiferromagnet.

Since the total Chern number of the gapped spin-$\uparrow$ and  
spin-$\downarrow$ spinons is zero, there is no Chern-Simons term for the 
gauge field, and thus the gauge field remains to be gapless. 
Usually the continuous lattice gauge theory without the gapless matter 
is expected to be confining, the observation and the argument 
of Ref.~\onlinecite{ran2009spontaneous} is that,
the U(1) gauge flux quantum carries the spin quantum number ${S^z=1}$. 
This is because the spin-$\uparrow$ spinon band
and the spin-$\downarrow$ spinon band carry the opposite Chern number ($\pm 1$).
The gauge flux quantum carries the spinon quantum number weighted by their Chern number. 
As the total $S^z$ is a conserved density even in the presence of the
out-of-plane DM interaction, the usual instanton events that cause the confinement 
are forbidden by the U(1) spin rotation symmetry, and the gapless mode of this conserved  
density implies that the global spin rotation symmetry around the $z$ direction 
is spontaneously broken in the massive Landau level state. 
Therefore, the gauge photon is nothing but the gapless Goldstone mode,
and the symmetry breaking is through the in-plane magnetic order.  
This in-plane magnetic order is compatible with the 
classical intuition that the out-of-plane DM interaction favors an in-plane
120-degree antiferromagnetic order. Unlike the degenerate case in Ref.~\onlinecite{ran2009spontaneous},
the direction of the out-of-plane DM vector selects the underlying spin chirality 
of the in-plane antiferromagnetic order. Moreover,
our theoretical result indeed suggests the 
{\sl increased} stability of the spinon Landau level state compared to the 
Fermi pocket state in the presence of the out-of-plane DM interaction.

With the antiferromagnetic order, there are the usual spin-wave-like magnon excitations,
where the gapless Goldstone mode is the gapless gauge photon in the dual language 
of the U(1) symmetric spin model. 
Both the gapless Goldstone mode and the usual spin-wave excitations are encoded in the 
$\langle S^+ S^-\rangle$ for the in-plane order. 
Moreover, we have the fractionalized spinon excitations that exhibit
continuous excitation spectra in the dynamic spin structure factor at the higher energies. 
A similar structure of magnetic excitations has been proposed
for the spin supersolid on the triangular lattice~\cite{Jia_2024,PhysRevB.111.104435,PhysRevLett.133.186704}.

Apart from the relevance with the kagom\'{e} lattice spin liquid in magnets, 
the flux generation picture could be readily applied to the 
recent proposal of kagom\'e Rydberg atom simulator with XY interactions~\cite{bintz2024diracspinliquidquantum}
where the Zeeman coupling is simply replaced by the detuning. Beyond that,
the spontaneous gauge flux generation for the gapless fermionic matter 
coupled to the U(1) lattice gauge theory is an interesting subject on its own. 
The system gains the kinetic energy for the fermions by spontaneously 
generating the U(1) gauge flux. 
For example, this may resolve the issue of the scalar spin chirality term 
that was proposed to understand the thermal Hall transport in cuprates, 
though the phonon contribution was later proposed~\cite{Samajdar_2019,Grissonnanche_2019}. 
In that context, it was argued theoretically that, 
the deconfined quantum criticality between the N\'{e}el and 
the valence bond state is dual to the gapless Dirac fermion coupled to dynamical U(1) gauge field. 
This spontaneous gauge flux generation by the magnetic field 
may naturally provide the scalar spin chirality term, though
the scalar spin chirality term is not independent and is related to the Zeeman coupling. 
More broadly, one could extend the spontaneous flux generation mechanism 
to many other matter-gauge
coupled systems. There are a couple more aspects for this extension. 
First, the gauge field does not really have to be continuous, 
though the change of the gauge flux 
is discrete. As we have mentioned in 
Sec.~\ref{sec:1}, Ref.~\onlinecite{Gazit_2017} has shown numerically for a $\mathbb{Z}_2$ gauge
theory that, the system shifts from the zero-flux state to the $\pi$-flux state. 
Second, the matter field does not have to be gapless. For the gapped matter fields,
the gauge field without the Chern-Simons term should be discrete to avoid the confinement. 
The kinetic energy gain of the matter field is through the filled Fermi sea 
for the fermionic matter and/or the quantum zero-point energy for the bosonic matter. 
One could in principle design concrete lattice gauge theory models to demonstrate this aspect
of physics.

To summarize, we have shown that, with the Dzyaloshinskii–Moriya interaction
and the weak Zeeman coupling for the kagom\'e lattice U(1) Dirac spin liquid, 
both the induced gauge flux and the spontaneous gauge flux are generated.
The properties of the resulting massive Landau level state is discussed.

\section*{Acknowledgments}

SYP and JHY acknowledge Pengwei Zhao, Ruitao Xiao, Zhengyu Xiao, Lingxian Kong and Yeyang Zhang for discussion,
and GC acknowledges Xiao-Gang Wen, Patrick Lee, Ying Ran and Shenghan Jiang for conversation. 
This work is supported by NSFC with Grants No.92565110 and No.12574061,
by the Ministry of Science and Technology of China
with Grants No.~2021YFA1400300, and by the Fundamental Research Funds for the Central Universities, 
Peking University.

\appendix

\section{Renormalized mean field theory}
\label{App:RMFT}
\setcounter{equation}{0}
\renewcommand{\theequation}{A.\arabic{equation}}
\renewcommand{\thesubsection}{\thesection.\arabic{subsection}}

In this section, we will deduce and present the form of the Gutzwiller factors $g_{zz}$ and $g_{+-}$ of the kagom\'{e} antiferro-magnetism. 

The renormalized mean field theory (RMFT) is used to calculate the expectation values of observables 
in projected mean-field states from a mean-field perspective. Different from the variational Monte-Carlo simulation (VMC), RMFT treats the Gutzwiller projection operator at the mean-field level via Wick's theorem rather than explicitly sampling configurations from the Hilbert space that meet physical constraints. 
In our work, the parton decomposition of spins requires the spinons to satisfy the single-occupancy constraint. 
Hence, the projection operator under half filling is defined as:
\begin{equation}\label{eq:proj_oper}
    \hat P_G\equiv \prod_{i}\hat P^i_G\equiv\prod_{i}\left[\hat n_{i\uparrow}(1-\hat n_{i\downarrow})+\hat n_{i\downarrow}(1-\hat n_{i\uparrow})\right],
\end{equation}
where $i$ is the index of the lattice sites. To calculate the Gutzwiller factor in Eq.~\eqref{eq:Gutz_factor} through RMFT, we need to compare the ratio of expectations of the operator $\hat O$ under $\ket{\psi}$ and $\ket{\psi_0}$, where $\ket{\psi_0}$ is the the mean-field ground state and $\ket{\Psi}=\hat P_{G}\ket{\Psi_0}$ is the physical state after Gutzwiller projection. For a local operator $\hat O$, $\braket{\hat O}_0\equiv\bra{\psi_0}\hat O\ket{\psi_0}$ can be obtained through Wick's theorem, so we subsequently focus on computing $\braket{\hat O}\equiv \bra{\psi}\hat O\ket{\psi}/\braket{\psi|\psi}$.

The total projection operator $\hat P_G$ can be decomposed into specific spin-configuration projection operators that satisfy the single-occupancy constraint:
\begin{equation}
    \hat P^{\{i_0\}}_{G}=\prod_{a\in\mathcal{A}_{\{i_0\}}}\hat n_{a\uparrow}(1-\hat n_{a\downarrow}
)\prod_{b\in\mathcal{B}_{\{i_0\}}}\hat n_{b\downarrow}(1-\hat n_{b\uparrow}),
\end{equation}
here $\{i_0\}$ is the index of the spin configurations that are single-spinon occupied per site. $\mathcal{A}_{i_0},\mathcal{B}_{i_0}$ is the set for the sites with spin-up and spin-down spinons of the $\{i_0\}$-th configuration.
Building upon the relation $\hat P_G=\sum_{\{i_0\}}\hat P^{\{i_0\}}_G$, following the work by Ogata and Himeda \cite{ogata2003superconductivity}, considering the single-occupancy constraint, the expectation of the operator can be calculated as 
\begin{equation}\label{eq:O_exp}
    \braket{\hat O}= \frac{\sum_{\{i_0\}}\braket{\hat O}_{\{i_0\}}}{W_{0}},
\end{equation}
where $\braket{\hat O}_{\{i_0\}}\equiv \bra{\psi_0}\hat P^{\{i_0\}}_G\hat O\hat P^{\{i_0\}}_G\ket{\psi_0}$. $W_0$ is defined as  
\begin{equation}\label{eq:W0}
    W_0=\sum_{\{i_0\}}\bra{\Psi_0}\hat P_{G}^{\{i_0\}}\ket{\Psi_0},
\end{equation}
which is identical for all operators. We will separately calculate $W_0$ and the numerator parts of $g_{zz}$ and $g_{+-}$.

Before proceeding with the actual calculations, we note that the Gutzwiller factors $g_{zz}$ and $g_{+-}$ depend on the bond order $\chi_{ij}^{\sigma}$ and the spin moment $m$ defined as:
\begin{equation}
    \left\{
\begin{aligned}
    &\chi_{ij}^{\sigma} \equiv\braket{f^\dagger_{i\sigma}f_{j\sigma}}_{0},\\
    &n \equiv\frac{N_S}{N}=\braket{\hat n_{ \uparrow}}_{0}+\braket{\hat n_{\downarrow}}_{0}, \\
    &m \equiv\frac{n_{\uparrow}-n_{\downarrow}}{2},\\
    &\braket{\hat n_{\uparrow}}_{0} =\frac{n}{2}+m\equiv r,\\
    &\braket{\hat n_{\downarrow}}_{0}  =\frac{n}{2}-m\equiv w,
\end{aligned}\right.
\end{equation}
where $\braket{}_0$ is the expectation under the normalized mean-field ground state $\ket{\psi_0}$, 
$\sigma$ is the index of spin, $i$ the index of the lattice sites. 
$N_S$ ($N$) is the total number of spinons (sites). 
To make the formula more concise, 
we assume that the modulus of $\chi^\sigma_{ij}$ 
is uniformly distributed on the nearest neighboring
bonds with
\begin{equation}
    \chi^\sigma_{ij}\equiv \chi^{\sigma}e^{i\phi_{ij^\sigma}}.
\end{equation}

\begin{figure}
    \centering
    \includegraphics[width=0.7\linewidth]{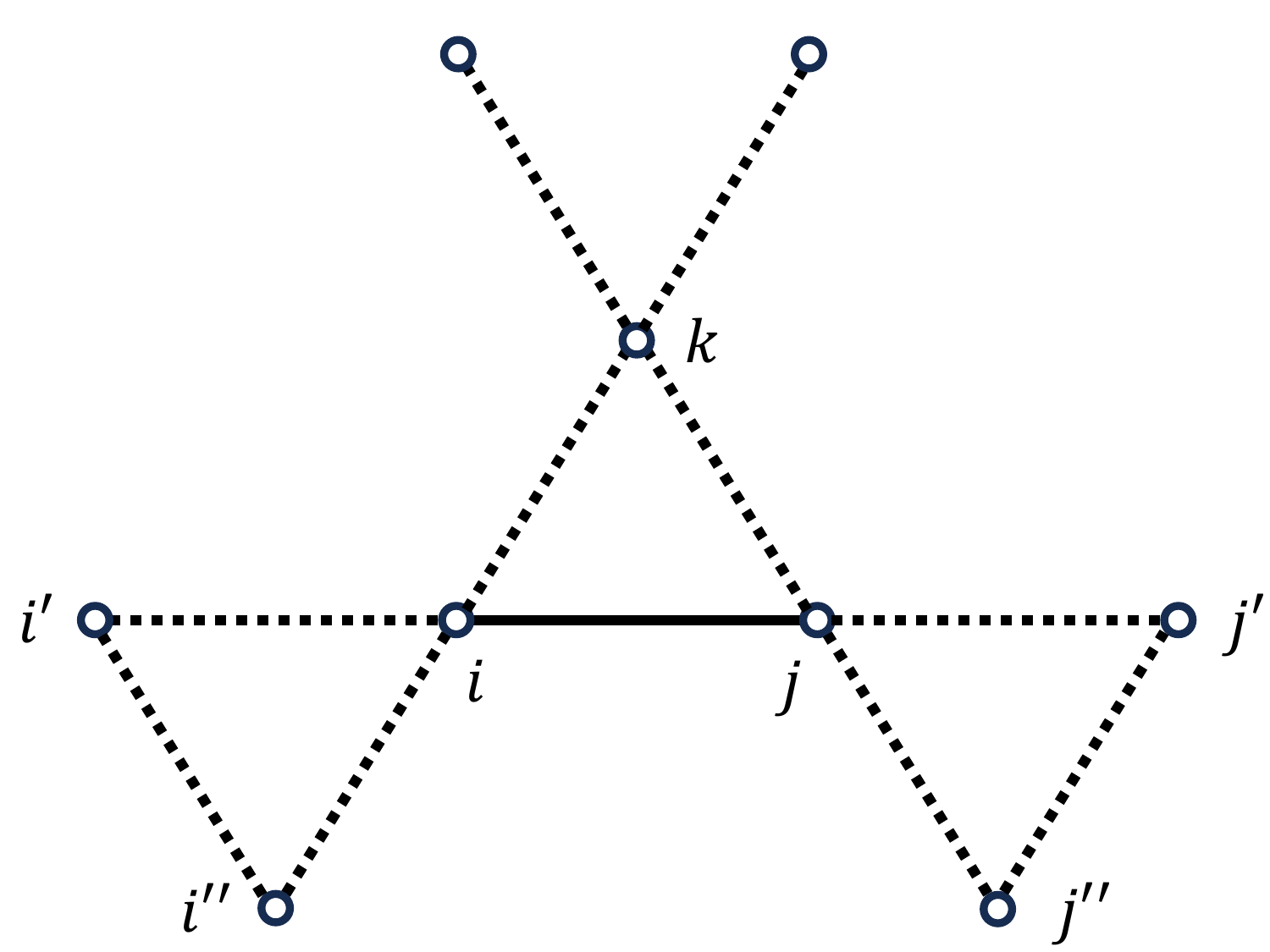}
    \caption{Lattice site index schematic diagram. $i,j$ are the sites of $S^\mu_iS^\nu_j$. $k$ is the nearest neighbor to both $i$ and $j$. $\{i^\prime, i^{\prime\prime},j^\prime, j^{\prime\prime}\}$ are the nearest neighbor to only $i$ or $j$. } 
    \label{fig:sites}
\end{figure}

\subsection{$W_0$}

$W_0$ defined in Eq.~\eqref{eq:W0} essentially evaluates $\braket{\hat P_G}_0=\braket{\prod_i\hat P_G^i}_0$. Through the Wick theorem, $W_0$ equals to the sum of the products of all the possible contraction strings. To achieve concise results, we only consider the contractions within the same site or the nearest neighbors. Furthermore, to preserve information of kagom\'{e} lattice, the connected diagram between three sites on one triangle plaquette will be taken into account. 

The lowest order is the site-diagonal expectation
\begin{equation}
    \braket{\hat P_G^i}_{0}=r(1-w)+w(1-r)=n-2rw.
\end{equation}

The next order will consider the contractions between the two nearest neighbor sites $\braket{\hat P_G^i\hat P_G^j}_{0}$. For $(i,j)$, there are four kinds of configurations $(\uparrow,\uparrow),(\uparrow,\downarrow),(\downarrow,\uparrow),(\downarrow,\downarrow)$, and we can calculate the expectation of all channels for these four configurations separately.
\begin{equation}
\begin{aligned}
P_{\uparrow\uparrow} & =\braket{n_{i\uparrow}(1-\hat n_{i\downarrow})n_{j\uparrow}(1-\hat n_{j\downarrow})}_{0}\\
& =r^{2}(1-w)^{2}-r^{2}|\chi^{\downarrow}_{ij}|^{2}-(1-w)^{2}|\chi^{\uparrow}_{ij}|^{2}+|\chi^{\uparrow}_{ij}|^{2}|\chi^{\downarrow}_{ij}|^{2},
\end{aligned}
\end{equation}

\begin{equation}
    \begin{aligned}
P_{\downarrow\downarrow} & =\braket{n_{i\downarrow}(1-\hat n_{i\uparrow})n_{j\downarrow}(1-\hat n_{j\uparrow})}_{0}\\
& =w^{2}(1-r)^{2}-w^{2}|\chi^{\uparrow}_{ij}|^{2}-(1-r)^{2}|\chi^{\downarrow}_{ij}|^{2}+|\chi^{\uparrow}_{ij}|^{2}|\chi^{\downarrow}_{ij}|^{2},
\end{aligned}
\end{equation}

\begin{equation}
    \begin{aligned}
P_{\uparrow\downarrow} & =\braket{n_{i\uparrow}(1-\hat n_{i\downarrow})n_{j\downarrow}(1-\hat n_{j\uparrow})}_{0}\\
& =r(1-w)w(1-r)+r(1-r)|\chi^{\downarrow}_{ij}|^{2}\\
&\quad +(1-w)w|\chi^{\uparrow}_{ij}|^{2}+|\chi^{\uparrow}_{ij}|^{2}|\chi^{\downarrow}_{ij}|^{2},
\end{aligned}
\end{equation}

\begin{equation}
\begin{aligned}
P_{\downarrow\uparrow} & =\braket{n_{i\downarrow}(1-\hat n_{i\uparrow})n_{j\uparrow}(1-\hat n_{j\downarrow})}_{0}\\
 & =w(1-r)r(1-w)+r(1-r)|\chi^{\downarrow}_{ij}|^{2}\\
 &\quad +(1-w)w|\chi^{\uparrow}_{ij}|^{2}+|\chi^{\uparrow}_{ij}|^{2}|\chi^{\downarrow}_{ij}|^{2}.
\end{aligned}
\end{equation}
By adding all the above terms together, $\braket{\hat P_G^i\hat P_G^j}_{0}$ is obtained as
\begin{equation}
    \begin{aligned}
        P_{\uparrow\uparrow}+ P_{\downarrow\downarrow}+P_{\uparrow\downarrow}+P_{\downarrow\uparrow}=(n-2rw)^{2}+X ,
    \end{aligned}
\end{equation}
where $X$ is the connected part within $i,j$ sites:
\begin{equation}\label{eq:X}
X=-4m^{2}|\chi^{\downarrow}_{ij}|^{2}-4m^{2}|\chi^{\uparrow}_{ij}|^{2}+4|\chi^{\uparrow}_{ij}|^{2}|\chi^{\downarrow}_{ij}|^{2}
\end{equation}

The highest order that 
we take into account is the connected part between the three sites $i,j,k$ on a triangular plaquette shown in Fig.~\ref{fig:sites}.
There are 8 configurations $(\uparrow\uparrow\uparrow),(\uparrow\uparrow\downarrow),(\uparrow\downarrow\uparrow
), (\downarrow\uparrow\uparrow), (\downarrow\downarrow\downarrow),(\downarrow\downarrow
\uparrow),(\downarrow\uparrow\downarrow), (\uparrow\downarrow\downarrow)$. We can obtain
\begin{equation}
    \begin{aligned}
    P_{\uparrow\uparrow\uparrow} & =\braket{n_{i\uparrow}(1-n_{i\downarrow})n_{j\uparrow}(1-n_{j\downarrow})n_{k\uparrow}(1-n_{k\downarrow})}_{0}\\
	& =\left[r^{3}-r(|\chi^{\uparrow}_{jk}|^{2}+|\chi^{\uparrow}_{ik}|^{2}+|\chi^{\uparrow}_{ij}|^{2})+( \chi^{\uparrow}_{ij}\chi^{\uparrow}_{jk}\chi^{\uparrow}_{ki}+h.c.)\right]                           \\
	& \times \left[(1-w)^{3}-(1-w)(|\chi^{\downarrow}_{jk}|^{2}+|\chi^{\downarrow}_{ik}|^{2}+|\chi^{\downarrow}_{ij}|^{2})\right.\\
    &\left. \quad \quad -( \chi^{\downarrow}_{ij}\chi^{\downarrow}_{jk}\chi^{\downarrow}_{ki}+h.c.)\right] ,
\end{aligned}
\end{equation}

\begin{equation}
    \begin{aligned}
	P_{\uparrow\uparrow\downarrow} & =\braket{n_{i\uparrow}(1-n_{i\downarrow})n_{j\uparrow}(1-n_{j\downarrow})n_{k\downarrow}(1-n_{k\uparrow})}_{0}\\
     & =\left[r^{2}(1-r)+r(|\chi^{\uparrow}_{jk}|^{2}+|\chi^{\uparrow}_{ik}|^{2})-(1-r)|\chi^{\uparrow}_{ij}|^{2}\right.\\
     &\quad \quad\left.-( \chi^{\uparrow}_{ij}\chi^{\uparrow}_{jk}\chi^{\uparrow}_{ki}+h.c.)\right]\\
	& \times \left[(1-w)^{2}w-w|\chi^{\downarrow}_{ij}|^{2}+(1-w)(|\chi^{\downarrow}_{jk}|^{2}+|\chi^{\downarrow}_{ik}|^{2})\right.\\
    &\left.\quad\quad +( \chi^{\downarrow}_{ij}\chi^{\downarrow}_{jk}\chi^{\downarrow}_{ki}+h.c.)\right]. \\
\end{aligned}
\end{equation}

The rest can be obtained by exchanging $\uparrow,\downarrow$ and $r,w$.  
For example $P_{\downarrow\downarrow\downarrow}=P_{\uparrow\uparrow\uparrow}(\uparrow \iff \downarrow,r\iff w )$, $P_{\uparrow\downarrow\uparrow}=P_{\uparrow\uparrow\downarrow}(j \iff k)$. 
Summing all eight cases and subtracting the contributions from the single-site connected term $(n-2rw)^{3}$ and two-site connected terms $3(n-2rw)X$, we can get the three-point connected part:
\begin{equation}
\label{eq:X1}
    \begin{aligned}
	X_{1}= & (P_{\uparrow\uparrow\uparrow}+P_{\uparrow\uparrow\downarrow}+...)-(n-2rw)^{3}-3(n-2rw)X \\
	=      & 16m^{3}(\chi^{\uparrow 3}\cos(\phi^{\uparrow}_{ijk})-\chi^{\downarrow 3}\cos(\phi^{\downarrow}_{ijk}))+48m^{2}\chi^{\uparrow 2}\chi^{\downarrow 2} \\
	       & -48m\chi^{\uparrow 2}\chi^{\downarrow 2}(\chi^{\uparrow }\cos(\phi^{\uparrow}_{ijk})-\chi^{\downarrow }\cos(\phi^{\downarrow}_{ijk}))              \\
	       & -32\chi^{\uparrow 3}\chi^{\downarrow 3}\cos(\phi^{\uparrow}_{ijk})\cos(\phi^{\downarrow}_{ijk})
\end{aligned}\,
\end{equation}
here $\phi_{ijk}^\sigma$ is defined as
\begin{equation}
    \phi^{\sigma}_{ijk}\equiv \phi^{\sigma}_{ij}+\phi^{\sigma}_{jk}+\phi^{\sigma}_{ki}.
\end{equation}

For a cluster with $N_{C}$ sites, there are totally $N_B$ bonds and $2/3N_C$ triangular plaquettes. If we select $l$ triangular contributions of $X_1$, $m$ bond contributions of $X$, the remaining contributions are $n-2rw$. To obtain $W_0$, we need to calculate the contribution of all the configurations, and then we can get
\begin{equation}
    \begin{aligned}
	W_{0}= & \sum_{l=0}^{2/3N_C}C_{\frac{2}{3}N_C}^{l}\sum_{m=0}^{N_B-9l}C_{N_B-9l}^{m}X_{1}^{l}X^{m}(n-2rw)^{N_C-3l-2m} \\
	= & a^{N_B}b^{\frac{2}{3}N_C}(n-2rw)^{N_C}.
\end{aligned}
\end{equation}
$C_N^M$ is the binomial coefficient. $a,b$ are given by
\begin{equation}\label{eq:ab}
    \begin{aligned}
	a=1+\frac{X}{(n-2rw)^{2}},\\
	b=1+\frac{X_{1}a^{-9}}{(n-2rw)^{3}}.
\end{aligned}
\end{equation}
We need to notice that the disconnected part in $\braket{\hat P_G^i\hat P_G^j}_0$ and $\braket{\hat P_G^i\hat P_G^j\hat P_G^k}_0$ has been calculated in the configuration which picks $l-1$ triangles to contribute $X_1$ and $m-1$ bonds to contribute $X$. When we pick one triangle to contribute $X_1$, then the 9 bonds connected to this triangle will not contribute $X$. This is why the summation over $m$ ranges from 0 to $N_B-9l$.

\subsection{$g_{+-}$}

With $W_0$, we need to calculate the numerator of Eq.~\eqref{eq:O_exp}. 
For $S^{+}_{i}S^{-}_{j}$, we need to calculate 
${\braket{S^{+}_{i}S^{-}_{j}}_{\{n_0\}}}=\braket{S^{+}_{i}S^{-}_{j}\hat P^{\{n_0\}}_G}_{0}$ 
where $\{n_0\}$ is used to label the configuration of the rest $N_{C}-2$ sites of the cluster other than $i,j$. 
As the same with $W_0$, the lowest order is given as 
\begin{equation}
    \braket{S^+_iS^-_j}_{0}(n-2rw)^{N_C-2}.
\end{equation}

Similar to $W_{0}$ if we select $l$ triangle contributions of $X_1$, $m$ bond contributions of $X$, 
the remaining contributions are $n-2rw$, we can get:
\begin{equation}
    \begin{aligned}
    \label{eq:SPSM_onsite}
     & \braket{S^+_iS^-_j}_{0}\sum_{l=0}^{2/3N_C-3}C_{\frac{2}{3}N_C-1}^{l}
       \sum_{m=0}^{\tilde N_B-9l}C_{\tilde N_B-9l}^{m}
       \\
     &\times X_{1}^{l}X^{m}(n-2rw)^{N_C-2-3l-2m}
       \\
     = & \braket{S^+_iS^-_j}_{0}a^{\tilde N_B}b^{\frac{2}{3}N_C-1}(n-2rw)^{N_C-2},
    \end{aligned}
\end{equation}
where $\tilde N_{B}$ is the number of links that are not connected to the $i,j$ sites. 
In the kagom\'e lattice, we have ${\tilde N_{B}=N_{B}-7}$.
Because $(i,j)$ will occupy $1$ bond and $3$ triangles, 
so the maximum number of possible triangles is reduced by $3$. 

Another correction to $g_{+-}$ involves considering the connected contraction between the local projection operator $\hat P^l_G $ and $S^{+}_{i}S^{-}_{j}$.  We only consider $l$ as the nearest neighbor of $i$ and $j$. For the kagom\'{e} lattice, there are two kinds of nearest neighbor sites of $i$ and $j$. The first kind is that $l$ is connected to only one of $i$ or $j$, whose connected contribution of $\braket{P^l_GS^+_iS^-_j}_{0}$ is zero. So we only consider the second kind, that $l$ is the nearest neighbor site $k$ of both $i,j$
sites. The connected part of $\braket{S^+_iS^-_j\hat P^k_G}_{0}$ is given by
\begin{equation}
\begin{aligned}
     X_2^{+-}&=\braket{S^+_iS^-_j\hat n_{k\uparrow}(1-\hat n_{k\downarrow})}_{c} +\braket{S^+_iS^-_j\hat n_{k\downarrow}(1-\hat n_{k\uparrow})}_{c}  \\
     &=\chi^{\downarrow}_{ji}\chi^{\uparrow}_{ik}\chi^{\uparrow}_{kj}(1-2w)+\chi^{\uparrow}_{ij}\chi^{\downarrow}_{ki}\chi^{\downarrow}_{jk}(1-2r)\\
     &+2\chi^{\uparrow}_{ik}\chi^{\downarrow}_{ki}\chi^{\uparrow}_{kj}\chi^{\downarrow}_{jk}.
\end{aligned}
\end{equation}
Similarly, the left $N_C-3$ sites will contribute
\begin{equation}
    \begin{aligned}\label{eq:X2pm}
	  & X^{+-}_{2}\sum_{l=0}^{2/3N_C-1}C_{\frac{2}{3}N_C-4}^{l}\sum_{m=0}^{\tilde N_B^{\prime}-9l}C_{\tilde N_B^{\prime}-9l}^{m}\\
      &\times X_{1}^{l}X^{m}(n-2rw)^{N_C-3-3l-2m} \\
	&=  X^{+-}_{2}a^{\tilde N_B^\prime}b^{\frac{2}{3}N_C-4}(n-2rw)^{N_C-3}.
\end{aligned}
\end{equation}
Here the number of the links that do not connect to the $i,j,k$ sites equals to
$    \tilde N_{B}^{\prime}=N_{B}-9$.

Combining $W_0$, Eq.~\eqref{eq:SPSM_onsite} and Eq.~\eqref{eq:X2pm} together, and taking into account that $\braket{S^+_iS^-_j}_{0}=\braket{f^\dagger_{i\uparrow}f_{i\downarrow}f^\dagger_{j\downarrow}f_{j\uparrow}}_{0}=-\chi^{\uparrow}_{ij}\chi^{\downarrow}_{ji}$, we can give $g_{+-}$ as
\begin{equation}\label{eq:gpm}
    \begin{aligned}
	g_{+-}=\frac{a^{-7}b^{-3}}{(n-2rw)^2}+\tilde X_{2}^{+-}\frac{a^{-9}b^{-4}}{(n-2rw)^3},
\end{aligned}
\end{equation}
where $a$, $b$, $X$, and $X_1$ has been defined in Eq.~\eqref{eq:ab},~\eqref{eq:X},~\eqref{eq:X1}. $\tilde X_2^{+-}\equiv {X_2^{+-}}/{\braket{S^+_iS^-_j}_0} $ is redefined as
\begin{equation}
    \begin{aligned}
   \tilde X_{2}^{+-} =&-\chi^{\uparrow}e^{-i\phi^{\uparrow}_{ijk}}(1-2w)-\chi^{\downarrow}e^{i\phi^\downarrow_{ijk}}(1-2r)\\
    &-2\chi^{\uparrow}\chi^{\downarrow}e^{i(\phi^\downarrow_{ijk}-\phi^\uparrow_{ijk})} 
    \end{aligned}.
\end{equation}

\subsection{$g_{zz}$}
For $S^{z}_{i}S^{z}_{j}$, there is some difference. As the same, the site-diagonal expectation, which is the lowest order contribution, is
\begin{equation}\label{eq:SZSZ_0}
    \braket{S_{i}^{z}S_{j}^{z}} _{0}(n-2rw)^{N_{C}-2}a^{\tilde{N}_{B}}
b^{2/3N_{C}-3}.
\end{equation}
Similar to $S^+_iS^-_j$, we need to consider the connected contraction from the nearest neighbor site $l$ to $i,j$. However, different from $S^+_iS^-_j$, when the nearest neighbor $l$ is only connected one of $i,j$, it can still contribute connected part to $\braket{S^z_iS^z_j}_{\{n_0\}}$. For example, $l$ is nearest neighbor only to $j$, so we have
\begin{equation}
    \begin{aligned}
	\braket{\hat S^z_i\hat S^z_jP^{l}_G}_{c} & =\braket{\hat S^z_i}_{0}\braket{\hat S^z_j\hat P^l_G}_{c}\\
	& =\frac{m}{2}\braket{ (\hat P^{j\uparrow}_{G}-\hat P^{j\downarrow}_{G} )(\hat P^{l\uparrow}_{G}+\hat P^{l\downarrow}_{G})}_{c} \\
	& = \frac{m}{2}(P_{\uparrow\uparrow}+P_{\uparrow\downarrow}-P_{\downarrow\uparrow}-P_{\downarrow\downarrow})_{c}\\
	 & =\frac{m}{2}(P_{\uparrow\uparrow}-P_{\downarrow\downarrow})_{c}\\
	 & =m^{2}(-|\chi^{\downarrow}_{jl}|^{2}-|\chi^{\uparrow}_{jl}|^{2}) \\
	& =m^{2}X_{3}.
\end{aligned}
\end{equation}
Here, we have omitted the $2m(n-rw)$ term in $P_{\uparrow\uparrow}-P_{\downarrow\downarrow}$ which is the disconnected part. We define $X_3$ as
\begin{equation}\label{eq:X3}
    X_{3}\equiv -|\chi^{\downarrow}_{jl}|^{2}-|\chi^{\uparrow}_{jl}|^{2}=-(\chi^{\uparrow})^{2}-(\chi^{\downarrow})^{2}.
\end{equation}

We use $N_m=2$ to denote the number of nearest neighbors unique to $i$, excluding the nearest neighbor $k$ site that is shared by both $i$ and $j$. Considering the contribution of the rest $N_C-3$ sites, if $l\in \{i^{\prime},i^{\prime\prime},j^{\prime},j^{\prime\prime}\}$, there will be $\tilde N_{B}^{\prime\prime}$ bonds which are not connected to $i,j,l$:
\begin{equation}
\tilde N_{B}^{\prime\prime}=N_{B}-10.
\end{equation}
Considering $i,j,l$ will occupy two triangles, so the total contribution of these parts will be:
\begin{equation}\label{eq:SZSZ_X3}
    2N_{m}m^{2}X_{3}(n-2rw)^{N_C-3}a^{\tilde N_B^{\prime\prime}}b^{2/3N_C-4}.
\end{equation}

But if $l=k$, the number of bonds that are not connected to $i,j,k$ will be $\tilde N_{B}^{\prime}$. Furthermore, when $l=k$ which is the nearest neighbor site of both $i$ and $j$ site, $\hat P_G^k$ can be connected to $i$ or $j$ or both $i$ and $j$ together. For the first case, the contribution is:
\begin{equation}
\braket{S^z_i}_{0}\braket{S^z_j\hat P_G^k}+\braket{S^z_j}_{0}\braket{S^z_i\hat P_G^k}
=2m^{2}X_{3}.
\end{equation}

For the second case, we use $\braket{}_c^\prime$ to label:
\begin{equation}
    \begin{aligned}
	\braket{S^z_iS^z_j\hat P^k_G}^{\prime}_{c} & =\braket{S^z_iS^z_j(\hat n_{k\uparrow}(1-\hat n_{k\downarrow})}^{\prime}_{c}\\
    &\quad +\braket{S^z_iS^z_j(\hat n_{k\downarrow}(1-\hat n_{k\uparrow})}^{\prime}_{c} \\
& =\frac{1}{4}\left[(\chi^{\uparrow}_{ij}\chi^{\uparrow}_{jk}\chi^{\uparrow}_{ki}+h.c.)(1-2w)\right. \\
&\quad +(\chi^{\downarrow}_{ij}\chi^{\downarrow}_{jk}\chi^{\downarrow}_{ki}+h.c.)(1-2r) \\
& \left.\quad +2\left(|\chi^{\uparrow}_{ik}|^{2}|\chi^{\downarrow}_{jk}|^{2}+|\chi^{\downarrow}_{ik}|^{2}|\chi^{\uparrow}_{jk}|^{2}\right)\right]\\
& =X_{3}^{\prime}.
\end{aligned}
\end{equation}
We define $X_3'$ as:
\begin{equation}\label{eq:X3P}
    X^{\prime}_{3}=m\left[\cos(\phi^{\uparrow}_{ijk})\chi^{\uparrow3}-\cos(\phi^{\downarrow}
_{ijk})\chi^{\downarrow3}\right]+\chi^{\uparrow2}\chi^{\downarrow2}.
\end{equation}
Combining the two terms, we obtain the one site connected channel contribution as
\begin{equation}\label{eq:SZSZ_X3X3P}
    (2m^{2}X_{3}+X_{3}^{\prime})(n-2rw)^{N_C-3}a^{\tilde N_B^\prime}b^{2/3N_C-4}
\end{equation}

Now we can evaluate the contributions of the two sites connected channel. The lowest order of the two sites connected channel is obvious:
\begin{equation}
    \braket{P^{m\prime}_GS^z_i}_{c}\braket{S^z_jP^{n\prime}_G}_{c}=m^{2}X_{3}^{2},
\end{equation}
here $m,n\in \{i^{\prime},i^{\prime\prime},j^{\prime},j^{\prime\prime},k\}$ shown in Fig.~\ref{fig:sites}. However, the choice of $m,n$ will decide the number of bonds that are not connected to $i,j,m,n$.

If $m\in \{i^{\prime},i^{\prime\prime}\},n\in\{j^{\prime},j^{\prime\prime}\}$, the number $\tilde N_{B}^{\prime\prime\prime }$ equals: 
\begin{equation}
    \tilde N_{B}^{\prime\prime\prime }=N_{B}-13.
\end{equation}
And there will be $N_m^2$ kinds of configurations of $\{m,n\}$, where $N_m=2$, so the total contribution of this case will be
\begin{equation}\label{eq:SZSZ_21}
    N_{m}^{2}m^{2}X_{3}^{2}(n-2rw)^{N_C-4}a^{\tilde N_B^{\prime\prime\prime}}b^{2/3N_C-5}.
\end{equation}

The second case is that one of $m,n$ is $k$ site, then the number of disconnected bond will be  $\tilde N_{B}^{\prime\prime\prime\prime}$. So the total contribution will be:
\begin{equation}\label{eq:SZSZ_22}
    2N_{m}m^{2}X_{3}^{2}(n-2rw)^{N_C-4}a^{\tilde N_B^{\prime\prime\prime\prime}}b^{2/3N_C-5}
\end{equation}

Combining $W_0$, Eq.~\eqref{eq:SZSZ_0},~\eqref{eq:SZSZ_X3},~\eqref{eq:SZSZ_X3X3P},~\eqref{eq:SZSZ_21},~\eqref{eq:SZSZ_22}, and taking into account that $\braket{S^z_iS^z_j}_{0}=m^{2}-\frac{1}{4}(\chi^{\uparrow2}+\chi^{\downarrow2})$, we can give $g_{zz}$ as 
\begin{equation}\label{eq:gzz}
    \begin{aligned}
	g_{zz}=\frac{a^{-7}b^{-3}}{(n-2rw)^2} & \left[1+\frac{2N_{m}m^2X_3}{m^2-\frac{1}{4}{(}\chi^{\uparrow2}+\chi^{\downarrow2})}\frac{a^{-3}b^{-1}}{(n-2rw)}\right.\\
    &+\frac{2m^2X_3+X_3^{\prime}}{m^2-\frac{1}{4}{(}\chi^{\uparrow2}+\chi^{\downarrow2})}\frac{a^{-2}b^{-1}}{(n-2rw)}        \\
	                                      & +\frac{N_{m}^2m^2X_3^2}{m^2-\frac{1}{4}{(}\chi^{\uparrow2}+\chi^{\downarrow2})}\frac{a^{-6}b^{-2}}{(n-2rw)^2}\\
                                          &\left.+\frac{2N_{m}m^2X_3^2}{m^2-\frac{1}{4}{(}\chi^{\uparrow2}+\chi^{\downarrow2})}\frac{a^{-5}b^{-2}}{(n-2rw)^2}\right],
\end{aligned}
\end{equation}
where $N_m=2$, and $X_3,X_3^\prime$ are defined in Eq.~\eqref{eq:X3},~\eqref{eq:X3P}.
$g^{+-}$ and $g^{zz}$ are given in Eq.~\eqref{eq:gpm},\eqref{eq:gzz}. Since the spinon system must satisfy the single-spinon occupied constraint, all instances of $n$ in the above formulas are equal to 1. 

The method used in the work by Ogata and Himeda \cite{ogata2003superconductivity} actually corrects the expectation value of the operator $\hat O$ under the Gutzwiller projection by calculating the real-space correlation between the operator $\hat O$ and the local projection operator $\hat P_G^i\equiv \Sigma_\sigma \hat n_{i\sigma}(1-\hat n_{i\bar{\sigma}})$ through an order-by-order computation. Different from their work on the square lattice, we consider the three-site correlation correction on a triangular plaquette of the kagom\'{e} lattice, which reflects the strong frustration in the kagom\'{e} material and the influence of spin scalar chirality fluctuations on the ground state.

\bibstyle{apsrev-nourl}
\bibliography{reference}
\end{document}